\documentclass[twoside,twocolumn,9pt]{article}
\usepackage{extsizes}
\usepackage[super,sort&compress,comma]{natbib} 
\usepackage[version=3]{mhchem}
\usepackage[left=1.5cm, right=1.5cm, top=1.785cm, bottom=2.0cm]{geometry}
\usepackage{balance}
\usepackage{widetext}
\usepackage{times,mathptmx}
\usepackage{sectsty}
\usepackage{graphicx} 
\usepackage{lastpage}
\usepackage[format=plain,justification=raggedright,singlelinecheck=false,font={stretch=1.125,small,sf},labelfont=bf,labelsep=space]{caption}
\usepackage{float}
\usepackage{fancyhdr}
\usepackage{fnpos}
\usepackage[english]{babel}
\usepackage{array}
\usepackage{droidsans}
\usepackage{charter}
\usepackage[T1]{fontenc}
\usepackage[usenames,dvipsnames,table]{xcolor}
\usepackage{setspace}
\usepackage[compact]{titlesec}
\usepackage{amsmath,amssymb}
\usepackage{multirow}
\usepackage{subfigure}

\definecolor{cream}{RGB}{222,217,201}

\newcommand{\Pe}{\rm{Pe}}
\newcommand{\ehat}{\vecv{\hat{e}}}
\newcommand{\be}{\begin{equation}}
\newcommand{\ee}{\end{equation}}
\newcommand{\bea}{\begin{eqnarray}}
\newcommand{\eea}{\end{eqnarray}}

\newcommand{\bw}{\begin{widetext}}
\newcommand{\ew}{\end{widetext}}
\newcommand{\lae}{\stackrel{<}{\scriptstyle\sim}}
\newcommand{\gae}{\stackrel{>}{\scriptstyle\sim}}

\newcommand{\kB}{k_{\rm B}}

\newcommand{\tauo}{\tau_{\rm o}}
\newcommand{\tauc}{\tau_{\rm c}}
\newcommand{\Dr}{D_{\rm r}}
\newcommand{\Dt}{\tilde{D}}
\newcommand{\va}{v_{\rm a}}
\newcommand{\vat}{\tilde{v}_{\rm a}}
\newcommand{\vatt}{\tilde{\tilde{v}}_{\rm a}}

\newcommand{\vecv}[1]{\mathbf{{#1}}}

\begin{document}

\pagestyle{fancy}
\thispagestyle{plain}
\fancypagestyle{plain}{

\fancyhead[C]{\includegraphics[width=18.5cm]{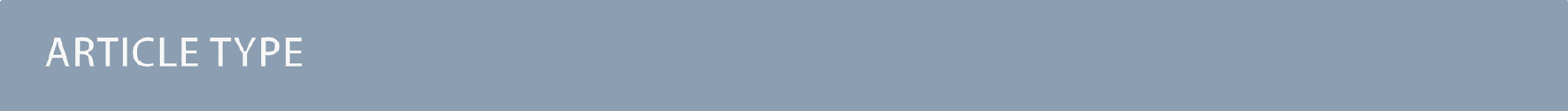}}
\fancyhead[L]{\hspace{0cm}\vspace{1.5cm}\includegraphics[height=30pt]{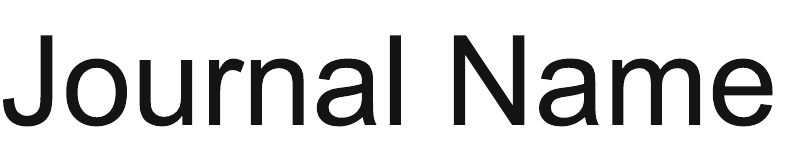}}
\fancyhead[R]{\hspace{0cm}\vspace{1.7cm}\includegraphics[height=55pt]{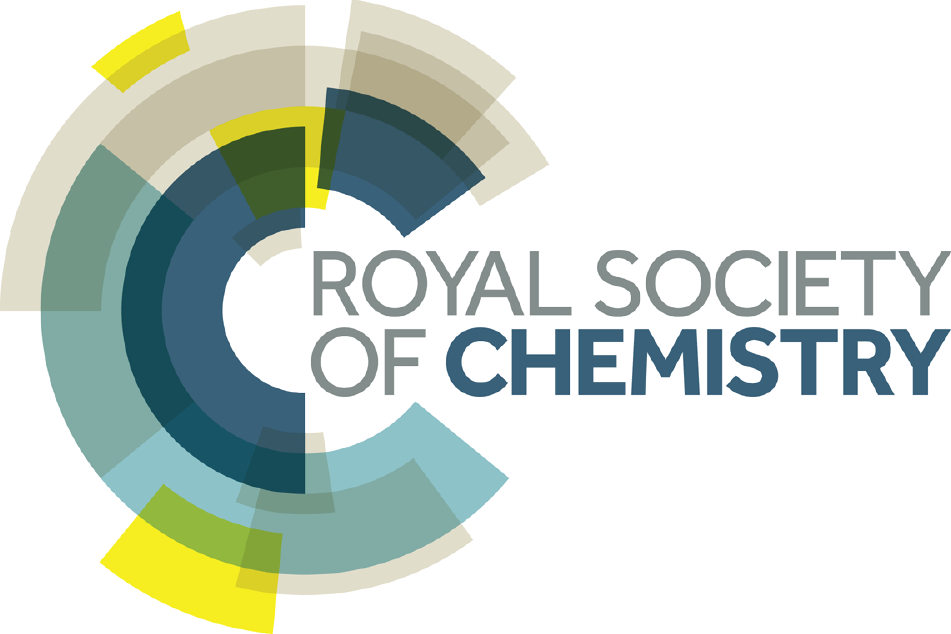}}
\renewcommand{\headrulewidth}{0pt}
}

\makeFNbottom
\makeatletter
\renewcommand\LARGE{\@setfontsize\LARGE{15pt}{17}}
\renewcommand\Large{\@setfontsize\Large{12pt}{14}}
\renewcommand\large{\@setfontsize\large{10pt}{12}}
\renewcommand\footnotesize{\@setfontsize\footnotesize{7pt}{10}}
\makeatother

\renewcommand{\thefootnote}{\fnsymbol{footnote}}
\renewcommand\footnoterule{\vspace*{1pt}%
\color{cream}\hrule width 3.5in height 0.4pt \color{black}\vspace*{5pt}} 
\setcounter{secnumdepth}{5}

\makeatletter 
\renewcommand\@biblabel[1]{#1}            
\renewcommand\@makefntext[1]%
{\noindent\makebox[0pt][r]{\@thefnmark\,}#1}
\makeatother 
\renewcommand{\figurename}{\small{Fig.}~}
\sectionfont{\sffamily\Large}
\subsectionfont{\normalsize}
\subsubsectionfont{\bf}
\setstretch{1.125} 
\setlength{\skip\footins}{0.8cm}
\setlength{\footnotesep}{0.25cm}
\setlength{\jot}{10pt}
\titlespacing*{\section}{0pt}{4pt}{4pt}
\titlespacing*{\subsection}{0pt}{15pt}{1pt}

\fancyfoot{}
\fancyfoot[LO,RE]{\vspace{-7.1pt}\includegraphics[height=9pt]{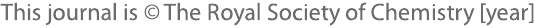}}
\fancyfoot[CO]{\vspace{-7.1pt}\hspace{13.2cm}\includegraphics{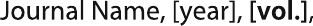}}
\fancyfoot[CE]{\vspace{-7.2pt}\hspace{-14.2cm}\includegraphics{head_foot/RF}}
\fancyfoot[RO]{\footnotesize{\sffamily{1--\pageref{LastPage} ~\textbar  \hspace{2pt}\thepage}}}
\fancyfoot[LE]{\footnotesize{\sffamily{\thepage~\textbar\hspace{3.45cm} 1--\pageref{LastPage}}}}
\fancyhead{}
\renewcommand{\headrulewidth}{0pt} 
\renewcommand{\footrulewidth}{0pt}
\setlength{\arrayrulewidth}{1pt}
\setlength{\columnsep}{6.5mm}
\setlength\bibsep{1pt}

\makeatletter 
\newlength{\figrulesep} 
\setlength{\figrulesep}{0.5\textfloatsep} 

\newcommand{\topfigrule}{\vspace*{-1pt}%
\noindent{\color{cream}\rule[-\figrulesep]{\columnwidth}{1.5pt}} }

\newcommand{\botfigrule}{\vspace*{-2pt}%
\noindent{\color{cream}\rule[\figrulesep]{\columnwidth}{1.5pt}} }

\newcommand{\dblfigrule}{\vspace*{-1pt}%
\noindent{\color{cream}\rule[-\figrulesep]{\textwidth}{1.5pt}} }

\makeatother

\twocolumn[
  \begin{@twocolumnfalse}
\vspace{3cm}
\sffamily
\begin{tabular}{m{4.5cm} p{13.5cm} }

\includegraphics{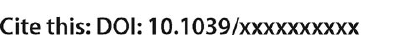} & \noindent\LARGE{\textbf{Clustering and phase behaviour of attractive active particles
  with hydrodynamics$^\dag$}} \\
\vspace{0.3cm} & \vspace{0.3cm} \\

 & \noindent\large{Ricard Matas Navarro,\textit{$^{a}$} and Suzanne
  Fielding$^{\ast}$\textit{$^{a}$}} \\

\includegraphics{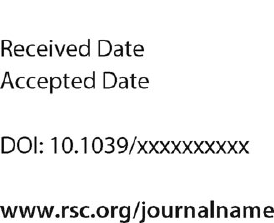} & \noindent\normalsize{We simulate clustering, phase separation and hexatic ordering in a
  monolayered suspension of active squirming disks subject to an
  attractive Lennard-Jones-like pairwise interaction potential, taking
  hydrodynamic interactions between the particles fully into account.
  By comparing the hydrodynamic case with counterpart simulations for
  passive and active Brownian particles, we elucidate the relative
  roles of self-propulsion, interparticle attraction, and hydrodynamic
  interactions in determining clustering and phase behaviour. Even in
  the presence of an attractive potential, we find that hydrodynamic
  interactions strongly suppress the motility induced phase separation
  that might a priori have been expected in a highly active
  suspension.  Instead, we find only a weak tendency for the particles
  to form stringlike clusters in this regime. At lower activities we
  demonstrate phase behaviour that is broadly equivalent to that of
  the counterpart passive system at low temperatures, characterized by
  regimes of gas-liquid, gas-solid and liquid-solid phase coexistence.
  In this way, we suggest that a dimensionless quantity representing
  the level of activity relative to the strength of attraction plays
  the role of something like an effective non-equilibrium temperature,
  counterpart to the (dimensionless) true thermodynamic temperature in
  the passive system. However there are also some important differences from the equilibrium case, most notably with regards the degree of hexatic ordering, which we discuss carefully.} \\

\end{tabular}

 \end{@twocolumnfalse} \vspace{0.6cm}

  ]

\renewcommand*\rmdefault{bch}\normalfont\upshape
\rmfamily
\section*{}
\vspace{-1cm}


\footnotetext{\textit{$^{a}$~Department of Physics, University of Durham,
  Science Laboratories, South Road, DH1 3LE, Durham, United Kingdom. E-mail: suzanne.fielding@durham.ac.uk}}

\footnotetext{\dag~Electronic Supplementary Information (ESI) available: movies of "living crystal" squirmer suspensions. See DOI: 10.1039/b000000x/}




\section{Introduction}
\label{sec:intro}

Active matter comprises internal mesoscopic subunits that collectively
drive the system far from thermal equilibrium by each individually
converting internal energy into mechanical work. In a biological
context, examples include crosslinked filaments activated by molecular
motors in the cell cytoskeleton~\cite{juelicher2007}, cells
collectively organised in living tissues~\cite{poujade2007},
suspensions of motile microorganisms~\cite{BergBook,rafai2010}, shoals
of fish and flocks of birds~\cite{ParrishHamnerBook}.  Synthetic
examples include vibrated granular
monolayers~\cite{narayan2007,Tsimring2008,deseigne2010} and
suspensions of phoretic colloidal
particles~\cite{PhysRevLett.108.268303,buttinoni13,PhysRevX.5.011004}.
As has been well documented in recent
reviews~\cite{ramas2010,RevModPhys.85.1143,Bialk2014}, active systems
display a host of exotic emergent effects including swarming, pattern
formation, giant number fluctuations and nonequilibrium ordering.
Among these, this paper focuses in particular on the phenomena of
aggregation and phase separation in active suspensions.

In the literature, (at least) two mechanisms have been put forward to
suggest that activity should generically lead to aggregation.  The
first~\cite{toner95,toner98,Toner2005170,ramastoner03,PhysRevE.77.011920,PhysRevLett.101.268101}
operates in systems of active particles that are elongated and so have
an intrinsic (steric) tendency to align with each other.  An
activity-mediated coupling between these orientational modes and
fluctuations in the local number density then provides a mechanism for
giant number fluctuations and phase separation. The standard deviation
$\Delta N$ of particle number in a subsystem with a mean number of
particles $N$ then scales as $N^a$, with $a > 1/2$.

The second mechanism~\cite{tailleur2008,cates2010,cates15} does not rely on
steric orientational effects and operates even in systems of
spherically symmetric particles. It was put forward originally in the
context of certain species of bacteria that obey simple
``run-and-tumble'' dynamics, in which each particle (at least when in isolation from any other particle) swims in straight lines at constant
speed between intermittent tumbling events in which it suddenly
changes swim direction.  A positive feedback loop then arises in which
the particles (a) accumulate in regions where they move more slowly
and (b) further slow down in regions where they are impeded by
crowding from other particles.  This renders an initially uniform
suspension unstable to ``motility induced phase separation''
(MIPS)~\cite{tailleur2008,cates2010}.  Following its prediction, MIPS
was indeed subsequently observed in simulations of spherical (or
disklike) active
particles~\cite{fily12,C3SM52469H,stenhammar14,redner13a}.

These simulations however discard hydrodynamic interactions between
the particles, which are of widespread importance in active systems.
Motivated by this, in Ref.~\cite{fielding14} we simulated a
suspension of active squirming disks with full hydrodynamics.  In
doing so we showed that hydrodynamic interactions in fact strongly
suppress MIPS, with the predicted phase separation replaced by only a
very weak stringlike clustering.

We interpreted this suppression in terms of the breakdown of a
mean-field assumption contained in the (a)-(b) feedback mechanism
outlined above: if a particle is to experience a slowing of its run by
crowding from other particles (part (b)), it must suffer many
collisions during any such run.  This assumption breaks down in the
case of hydrodynamic squirmers, because whenever two squirmers scatter
off each other hydrodynamically they each suffer an $O(1)$ change in
swim direction~\cite{scatter}.  To support this interpretation, we further
demonstrated suppression of MIPS even without hydrodynamics, in active
Brownian particles in any regime where the time for the decorrelation
of swim direction fails to comfortably exceed the time between
particle collisions.  The importance of rotational and translational
timescales in determining phase separation was also discussed in
Refs.~\cite{PhysRevE.89.062301,bialk2013}.

Nonetheless, strong aggregation effects have been widely observed
experimentally in active suspensions such as monolayered
light-activated colloidal surfers~\cite{palacci13} and phoretic
colloidal Janus
particles~\cite{PhysRevLett.108.268303,buttinoni13,PhysRevX.5.011004}.
Conversely, however, many bacterial suspensions show pronounced
coherent structures in their velocity field without strong spatial
inhomogeneities in
concentration~\cite{sokolov07,Shawnd13,PhysRevLett.110.228102},
although aggregation has been studied in suspensions of bacteria (with
and without active flagellae) subject to an attractive depletion
interaction~\cite{Schwarz-Linek13032012,C0SM00214C,Dorken07122012}.

In many of these experimental systems, the presence of a slight
attractive potential between the particles cannot be ruled out. In
contrast, the simulation study of Ref.~\cite{fielding14} concerned
particles that, besides hydrodynamic effects, interact only via a
purely repulsive potential that onsets steeply at particle-particle
contact.  The original predictions of MIPS likewise concerned purely
repulsive hard particles~\cite{tailleur2008,cates2010}.  A key open
question, therefore, is whether the presence of an attractive
contribution to the inter-particle potential might restore a tendency
for active suspensions to phase separate, even in the presence of
hydrodynamics.

As noted above, the phase behaviour of active particles has been
extensively studied in the absence of hydrodynamic interactions by
means of Brownian dynamics simulations, both
without~\cite{fily12,C3SM52469H,stenhammar14,PhysRevE.89.062301,bialk2013,bialk2012,PhysRevLett.113.028103,
  PhysRevLett.112.220602,wysocki2014,niran2013,C4SM01273A,soto14} and
with~\cite{PhysRevLett.111.245702,redner13b,prymidis15} attractive
interactions between the particles.  In particular,
Ref.~\cite{redner13b} reported attraction-dominated phase separation
at low activity, with a crossover to MIPS at higher levels of
activity.  However these existing studies of attractive active systems
again neglect hydrodynamic interactions. Given the strong effect of
hydrodynamics on the phase behaviour of purely repulsive active
suspensions~\cite{fielding14}, an important outstanding problem is to
elucidate the phase behaviour of attractive active suspensions with
hydrodynamics.

With that motivation in mind, in this work we simulate a suspension of
active squirming particles~\cite{light52,Blake} with an attractive
Lennard-Jones component to the potential of interaction between them.
Hydrodynamic interactions are taken fully into account using the
simulation method of Ref.~\cite{fielding14}, which allows for short
range lubrication, long range (power law) propagators, and indeed
hydrodynamic effects on all lengthscales in between.  The method
moreover gives access to the full range of packing fractions from
dilute to close packed, thereby allowing a comprehensive study of
phase behaviour. In these hydrodynamic simulations we neglect Brownian
motion, considering the limit of purely athermal dynamics that is
relevant to many active systems, such as bacterial suspensions.  We
return below to justify this neglect more fully by commenting on the
likely effect of a non-zero temperature.

To understand the effects of activity in the face of this attractive
component to the interaction potential, we define a dimensionless
parameter that quantifies the strength of swimming compared to the
strength of the attractive interaction.  For small values of this
parameter (low activity), we shall find phase behaviour that is
dominated by this attractive interaction, as might be expected
intuitively. In particular, we find regimes of phase coexistence
(gas-solid, gas-liquid, liquid-solid, {\it etc.}) that closely
resemble those of passive, thermalised LJ particles.  In this sense,
the scaled activity parameter for the active system might be viewed as
a dimensionless effective temperature that plays a role somewhat
analogous to the true dimensionless temperature in the counterpart
passive LJ system.  There are however some important differences
between the attraction-dominated phase separation that we report here
in the active system and its equilibrium counterpart, most notably
with regards the degree of crystalline ordering.

In contrast, for high levels of activity we find phase behaviour that
is relatively unaffected by the LJ attraction. This is also consistent
with intuition, because strongly swimming particles are expected to be
much less affected by the attractive interaction.  Indeed in this
regime we find only a weak tendency for stringlike clustering. We
give evidence to support the interpretation that this is again the
signature of a nearby MIPS that has been narrowly avoided by the
hydrodynamic suppression mechanism~\cite{fielding14} outlined
above for purely repulsive particles.

In between these two limiting regimes - attraction-dominated phase
separation at low activity, and stringlike clustering at high activity
- we find a slow crossover region in which the two effects merge, with
a degree of particle clustering that is set by the packing fraction
and the dimensionless activity parameter. 

Alongside the attractive Lennard Jones contribution to the interaction
potential, a steep repulsive interaction is also present that diverges
at particle contact, {\it i.e.,} at a particle separation $r\to 2R^+$,
where $R$ is the bare (hydrodynamic) radius of the squirmers. We
therefore expect much stronger excluded volume effects than for the LJ
potential that is conventionally used in the literature, which
diverges only at full overlap $r\to 0^+$.  In view of this, as a
warmup to our hydrodynamic simulations we shall first map out the
phase behaviour of both passive and active Brownian particles with
this modified LJ potential, in the absence of hydrodynamics. This will
then provide a benchmark against which the results of the hydrodynamic
study can be properly compared.

Because of the heavy computational cost of simulating dense
suspensions with full hydrodynamics, we have only been able to access
a relatively small number of particles in each run, typically $N=242$.
Phase diagrams will therefore be elucidated in qualitative outline
only, most obviously near any critical point.  Indeed this provides an
additional motivation for first simulating passive and active Brownian
particles, in order to learn what features of phase behaviour we can
realistically expect to capture with this small $N=242$. Accordingly,
although we could in principle achieve much larger $N=10^5$ or even
$N=10^6$ in the Brownian simulations, we use $N=242$ here too, to
provide a true comparison with the hydrodynamic results. While this
small $N$ is clearly a limitation, it is worth remarking that some of
the experiments reporting clustering of artificial swimmers also
involve a relatively small number of particles.  We note also that the
first simulations to map out the phase behaviour of passive Lennard
Jonesium used a similarly small
$N$~\cite{Binder1990,Hansen69,Alder1960}.

The paper is organized as as follows. In Sec.~\ref{sec:models} we
describe the Brownian and hydrodynamic models to be studied, along
with the simulation methods used. Sec.~\ref{sec:parameters} summarizes
the model parameters, our choice of units, and the principal
adimensional quantities to be explored in the simulations.  In
Sec.~\ref{sec:initial} we discuss the initial condition used in the
simulations, before in Sec.~\ref{sec:measures} defining the
statistical quantities that we shall measure during the runs.  In
Sec.~\ref{sec:results} we present our results for passive Brownian
particles, active Brownian particles and hydrodynamic squirmers in
turn.  Sec.~\ref{sec:conclusions} contains our conclusions.

\section{Models}
\label{sec:models}

In this section we introduce the three models to be studied in the
remainder of the manuscript. We start by defining the modified
Lennard-Jones interaction potential that is common to all three
models. We then define the dynamical rules of the three models
separately: passive Brownian particles (PBP), active Brownian
particles (ABP) and hydrodynamic squirmers (HS or simply squirmers).
Note that each model has its own set of dynamical rules, but all are
subject to the same modified Lennard-Jones potential.

For all three models we consider a two dimensional (2D) system
comprising $N$ disks in a square box of side $L$ with periodic
boundary conditions.  Each particle has a bare radius $R$ (bare in a
sense to be defined below), implying a bare area fraction $\phi_r=N\pi
R^2/L^2$.

\subsection{Potential}
\label{sec:potential}

\begin{figure}[h]
\centering
\includegraphics[width=8.0cm]{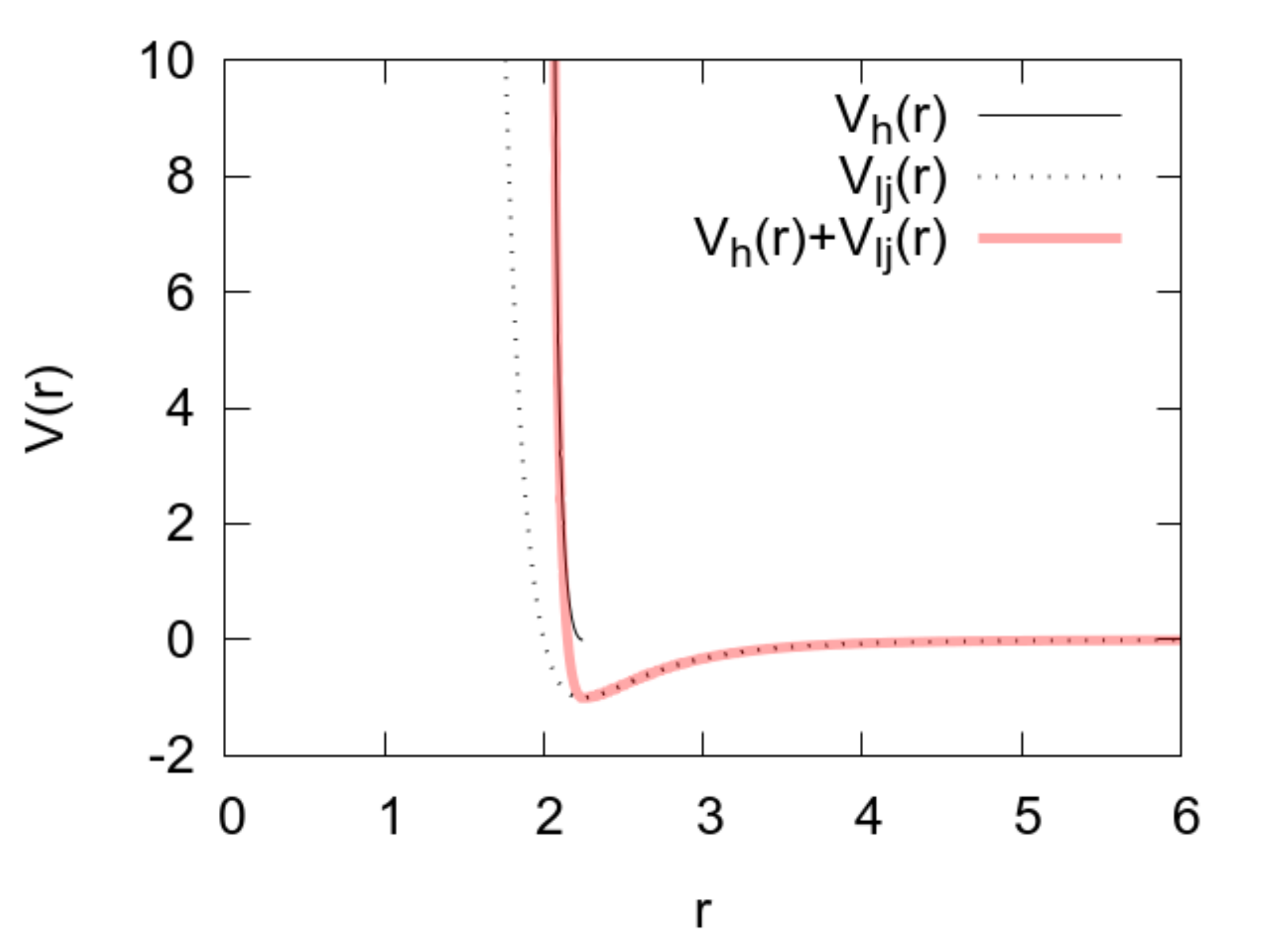}
\caption{\label{figpot}Modified Lennard-Jones potential in units of
  $\epsilon$=1, $R$=1. Lengthscales $\sigma_{\rm LJ}=2$ and
$\sigma_{\rm h}=2(2^{1/6}-1)$.}
  
\end{figure}

The particles experience pairwise Lennard-Jones interactions governed
by the potential
\begin{eqnarray}\label{ljpot}
V_{\rm LJ}(r)=4\epsilon_{\rm LJ}  \left[\left(\frac{\sigma_{\rm LJ}}{r}\right)^{12}- \left(\frac{\sigma_{\rm LJ}}{r}\right)^{6}\right],
\end{eqnarray}
which gives short-ranged repulsion for particle separations
$r<2^{1/6}\sigma_{\rm LJ}$, and long-ranged attraction for particle
separations $r>2^{1/6}\sigma_{\rm LJ}$.  See the dotted line in
Fig.~\ref{figpot}.  Setting $\sigma_{\rm LJ}=2R$ then sets the minimum
of $V_{\rm LJ}$ to be located at $r=2^{7/6}R$.

As we shall describe in more detail below, the hydrodynamically
squirming disks propel themselves by effecting a tangential velocity
round their edges, at a hydrodynamic radius $R$ from the particle
centres. Given the hydrodynamic interactions that we shall also
describe, short range lubrication effects then in principle disallow
any particle overlaps in which particles approach each other to a
separation $r<2R$. However the inevitably finite discretisation of
time and space in the simulation can in practice result in occasional
overlaps. To prevent these we follow standard practice in the
literature and include an additional short ranged steep repulsive
contribution
\begin{equation}
V_{\rm h}(r)=\epsilon_{\rm h}
\left[\left(\frac{\sigma_{\rm h}}{r-2R}\right)^{\nu}-\left(1-(r-2R)\frac{\nu}{\sigma_{\rm h}}\right)\right]
\end{equation}
for $2R\leq r\leq 2R+\sigma_{\rm h}$, with $V_{\rm
  h}(r)=0$ otherwise.  We then choose a shell size $\sigma_{\rm
  h}=(2^{1/6}-1)2R$ such that the onset of this steep repulsive
interaction at particle separations $r=2R+\sigma_{\rm h}$ coincides
with the minimum in the Lennard-Jones potential at
$r=2^{1/6}\sigma_{\rm LJ}=2^{7/6}R$. See the thin solid line in
Fig.~\ref{figpot}. The divergence of this additional contribution as
$r\to 2R^+$ then prohibits overlaps in which squirmers approach each
other to a separation of less than twice their hydrodynamic radius.

For simplicity we set the potential amplitudes equal throughout,
$\epsilon_{\rm LJ}=\epsilon_{\rm h}\equiv \epsilon$, and use a value
for the exponent $\nu=2$. The final combined pairwise interaction
potential $V=V_{\rm LJ}+V_{h}$ is shown by the thicker solid line in
Fig.~\ref{figpot}. Although the additional repulsive shell provided by
$V_{\rm h}$ is not needed for the PBP or ABP (recall that we include
it simply to avoid occasional hydrodynamic overlaps in the squirmers),
we use it for the Brownian models too to ensure a consistent potential
in all simulations, thereby allowing a truly direct comparison between
all three models.

The presence of this steeply repulsive shell acts to confer an
increased effective disk radius $R_{eff}=R+\sigma_{\rm h}/2$ that
slightly exceeds the bare disk radius $R$. Accordingly we define an
effective area fraction $\phi=N\pi R_{eff}^2/L^2$ that likewise
slightly exceeds the bare one. Whenever we quote an area fraction
throughout the rest of the paper, it is this effective quantity
$\phi$.

\subsection{Dynamics}
\label{sec:dynamics}

\subsubsection{Passive Brownian particles (PBP).}
\label{sec:PBP}

Before considering active suspensions, which are the primary focus of
this work, we will first consider passive Brownian particles (PBP)
subject to the same modified Lennard-Jones potential discussed above.
We shall do so in order first to establish the equilibrium phase
behaviour of a system of particles with this modified potential, which
will then serve as a point of reference for studying the
non-equilibrium phase behaviour of active systems subject to the same
potential, both without (ABP) and with (HS) hydrodynamic interactions.

Within the model of passive Brownian dynamics, the position
$\vecv{r}_i$ of the $i$th particle evolves according to
\begin{eqnarray}
\frac{d\vecv{r}_i}{dt}= -\frac{1}{\gamma}\sum_{j\neq i}\frac{\partial V(|\vecv{r}_i-\vecv{r}_j|)}{\partial \vecv{r}_{i}} + \sqrt{2 D} \Xi_i,
\end{eqnarray}
with the effect of a solvent coarse-grained in terms of an effective
drag of coefficient $\gamma$, and with a thermal noise characterised
by a diffusion coefficient $D$. The Gaussian random variable
$\Xi$ has
\begin{eqnarray}
  <\Xi_{i,\alpha}(t)>&=&0,\nonumber\\
  <\Xi(t)_{i,\alpha}\Xi_{j,\beta}(t')>&=&\delta_{ij}\delta_{\alpha\beta}\delta(t-t'),
\end{eqnarray}
where Latin and Greek symbols denote particle number and Cartesian
coordinate respectively. The drag is related to the diffusion constant
by the Einstein relation $D=\kB T/\gamma$, where $T$ is the
thermodynamic temperature and $\kB$ is Boltzmann's constant.  The
particles also suffer stochastic rotation following
Eqn.~\ref{eqn:ABPdynamicsR} below, with $\Dr=3 D/ 4 R^2$. However this
is irrelevant, because the translational dynamics is independent of
particle orientation or rotation for these passive particles.

\subsubsection{Active Brownian particles (ABP).}
\label{sec:ABP}

Within the model of active Brownian particles, the translational
velocity of the $i$th particle is given by
\be
\label{eqn:ABPdynamicsT}
\frac{d \textbf{r}_i}{dt}=  -\frac{1}{\gamma}\sum_{j\neq i}\frac{\partial V(|\vecv{r}_i-\vecv{r}_j|)}{\partial \vecv{r}_{i}} + v_a \textbf{e}_i.
\ee
Besides the effect of the conservative force stemming from the
modified Lennard-Jones potential, each particle also moves with a
self-propelling velocity $v_a \textbf{e}_i$. The speed $v_a$ is an
imposed constant of the problem. The unit vector
$\textbf{e}_i=(\cos{\alpha_i},\sin{\alpha_i})$ prescribes the
instantaneous preferred direction of self-propulsion, and obeys
stochastic angular dynamics governed by a rotational diffusion
constant $\Dr$, such that
\be
\label{eqn:ABPdynamicsR}
\frac{d \alpha_i}{dt}=\sqrt{2\Dr}\Xi_i(t).
\ee
The Gaussian random variable $\Xi_i$ obeys
\begin{eqnarray}
  <\Xi_{i}(t)>&=&0,\nonumber\\
  <\Xi_{i}(t)\Xi_{j}(t')>&=&\delta_{ij}\delta(t-t').
\end{eqnarray}

Note that in an equilibrium system the rotational diffusion
coefficient $\Dr$ would have to relate to the corresponding
translational diffusion coefficient $D$ as $\Dr=3 D/ 4 R^2$.  However
the active Brownian particles of interest here do not need to follow
any such relation.  Indeed, the stochastic angular dynamics prescribed
above is not intended to be of true thermal original, but rather as a
continuous-time statistical model of (for example) tumbling events in
a system of run-and-tumble bacteria, or of particle reorientations
occurring via hydrodynamic scattering off other particles. The
equivalent rotational diffusion coefficient that characterises these
active phenomena is expected be much larger in magnitude than the true
thermal one. In order to focus on the effects of activity, therefore,
we have removed the thermal noise from the translational equation
altogether, setting the translational diffusion coefficient $D=0$
upfront~\cite{fily12,C3SM52469H}.  Indeed we suggest that use of the
terminology ``Brownian'' for ABPs is perhaps better avoided
altogether, though we keep it for consistency with the earlier
literature.

\subsubsection{Hydrodynamic squirmers (HS).}

The model of hydrodynamic squirmers is based on a minimal description
of microbial propulsion intended to mimic locomotion through a fluid
by the beating of many cilia on the surface of a microbe
~\cite{light52,Blake,Blake2D,ishi08,ishi08b}.  The particles are
assumed to be neutrally buoyant and so force free.  Their size and
swimming speed are sufficiently small that the Reynolds number is
assumed negligible (zero inertia) and their swimming dynamics purely
Stokesian. Their size is however large enough that Brownian motion
(both rotational and translational) is assumed negligible,
corresponding to an infinite P\'eclet number.  Other sources of
rotational noise such bacteria tumbling are also neglected, and we
return to justify this assumption carefully below.

Consider then a monodisperse suspension of $N$ inertialess disklike
swimmers, each of radius $R$, and each propelling itself through an
incompressible Newtonian fluid in the $x-y$ plane by means of an
active ``squirming'' motion modeled by a prescribed tangential
velocity
\be
\label{eqn:slip}
S(\theta-\alpha_i)=B_1\sin(\theta-\alpha_i)+\frac{1}{2}B_2\sin2(\theta-\alpha_i)
\ee
round the disk edge, at a distance $R$ from the disk's centre.  For
simplicity this tangential velocity is taken to be time-independent,
representing an average over many beating cycles.  We denote by
$\beta=B_2/B_1$ the ratio of the second and first modes. The sign of
$\beta$ then distinguishes between ``pushers'' and ``pullers'', with
positive $\beta$ corresponding to pullers and negative $\beta$ to
pushers.  The amplitude of $\beta$ prescribes the degree of polarity,
with $|\beta|\ll 1$ giving strongly polar swimmers and $|\beta|\gg 1$
apolar ``shakers''.

A single disk undisturbed by any others in an infinite box then has a
swim speed $v_0=B_1/2$~\cite{Blake2D}, with an instantaneous preferred
swim direction $\ehat_i=(\cos\alpha_i,\sin\alpha_i)$.  In a suspension
of many disks the actual swim speeds and directions evolve over time
due to hydrodynamic interactions mediated by the Newtonian fluid
through which the particles move, as we now describe.

The velocity and pressure fields $\vecv{v}(\vecv{r},t)$ and
$p(\vecv{r},t)$ in the fluid outside the disks obey the condition of
mass balance for incompressible flow
\be
\label{eqn:incomp}
0=\nabla \cdot \vecv{v}(\vecv{r},t),
\ee
and the condition of force balance in the inertialess limit
\be
\label{eqn:stokes}
0 = \eta \nabla^2\vecv{v}(\vecv{r},t) - \nabla p(\vecv{r},t)+\vecv{f}.
\ee 
Here $\eta$ is the fluid viscosity and $\vecv{r}=(x,y)$ is a position
vector in the $x-y$ plane.  (For convenience the interior of the discs
is also taken to contain Newtonian fluid obeying
Eqns.~\ref{eqn:incomp} and~\ref{eqn:stokes}.  We then simply discard
this part of the mathematical solution, because we are only interested in
the behaviour of the fluid outside the disks.)


\begin{table*}
  \centering
\begin{tabular*}{\textwidth}{@{\extracolsep{\fill}}|l|c|c|c|c|c|c|c|}
\hline
Quantity&Symbol&Dimensions&Dimensionless form &Value&PBP&ABP&HS\\
\hline
Particle radius&$R$&$L$&1 &length unit&\checkmark &\checkmark&\checkmark\\
Potential energy&$\epsilon$&$GL^3$&1 &modulus unit&\checkmark &\checkmark&\checkmark\\
Drag coefficient&$\gamma$&$GTL$&1 &time unit for PBP/ABP&\checkmark&\checkmark&$\times$\\
Solvent viscosity&$\eta$&$GT$&1 &time unit for HS&$\times$&$\times$&\checkmark\\
Number of particles&$N$&1&N&242&\checkmark&\checkmark&\checkmark\\ 
Potential shell size&$\sigma_{\rm h}$&$L$&$\tilde{\sigma}_{\rm h}=\frac{\sigma_{\rm h}}{R}$&$2(2^{1/6}-1)$&\checkmark&\checkmark&\checkmark\\
Lennard-Jones length &$\sigma_{\rm LJ}$&$L$&$\tilde{\sigma}_{\rm LJ}=\frac{\sigma_{\rm LJ}}{R}$&2&\checkmark&\checkmark&\checkmark\\
Effective area fraction&$\phi$&1&$\phi$&to be varied (all models)&\cellcolor[HTML]{E3E3E3}{\color[HTML]{000000}\checkmark }& {\cellcolor[HTML]{E3E3E3}\checkmark}&\cellcolor[HTML]{E3E3E3}{\color[HTML]{000000}\checkmark}\\ 
Thermal diffusion coefficient&$D$&$L^2T^{-1}$&$\tilde{D}=\frac{D\gamma}{\epsilon}$& to be varied (PBP) &\cellcolor[HTML]{E3E3E3}{\color[HTML]{000000}\checkmark}&$\times$&$\times$\\
Rotational diffusion coefficient&$\Dr$&$T^{-1}$&
$\zeta^{-1}=\frac{\Dr R}{v_{\rm a}}$&to be varied (ABP)&$-$&\cellcolor[HTML]{E3E3E3}{\color[HTML]{000000}\checkmark}&$\times$\\
Active swim speed&$v_a$&$LT^{-1}$&$\vat$&to be varied (ABP $+$ HS)&$\times$&\cellcolor[HTML]{E3E3E3}{\color[HTML]{000000}$\frac{v_a\gamma R}{\epsilon}$}&\cellcolor[HTML]{E3E3E3}{\color[HTML]{000000}$\frac{B_1\eta R^2}{2\epsilon}$}\\
Active polarity&$\beta$&1&$\beta$&to be varied (HS) &$\times$&$\times$&\cellcolor[HTML]{E3E3E3}{\color[HTML]{000000}\checkmark}\\
\hline
\end{tabular*}
\caption{\label{parameters} Parameters for passive Brownian particles (PBP), active Brownian particles (ABP) and hydrodynamic squirmers (HS)  with modified Lennard-Jones interactions. Note that PBP are rotationally symmetric so rotational diffusion is unimportant. We work in dimensions space $GLT$ of modulus $G$, length $L$ and time $T$, rather than the more usual $MLT$ with $M$ being mass. 
}
\end{table*}

We recognise Eqn.~\ref{eqn:stokes} as the Stokes equation subject to
additional source terms $\vecv{f}$. These represent forces round the
edge of each disk:
\be
\label{eqn:forces}
\vecv{f}(\vecv{r},t)=\sum_i\vecv{f}_i(\theta_i)\delta(r_i-R),
\ee
in which we are summing here over the separate polar coordinate
systems $(r_i,\theta_i)$ of the disks, such that for the $i$th disk
\be
\vecv{r}=\vecv{R}_i(t)+r_i\cos(\theta_i)\hat{\vecv{x}}+r_i\sin(\theta_i)\hat{\vecv{y}},
\ee
where $\vecv{R}_i(t)$ is the position of the disk's centre. 

These forces are included so as to ensure that the $i$th disk has at
any instant a velocity round its edge
\be
\label{eqn:velocities}
\vecv{v}_i=\vecv{V}_i - R\Omega_i\hat{\vecv{\theta}}_i +
S(\theta-\alpha_i)\hat{\vecv{\theta}}_i,
\ee
comprising solid body translation and rotation, plus the
tangential squirming motion prescribed by the slip velocity function
in Eqn.~\ref{eqn:slip} above.

At any timestep $t \to t + Dt$ in the simulation, given a
current set of particle positions $\vecv{R}_i$ and preferred swim
directions $\alpha_i$, known quantities are the lowest two modes of
the force (which are constrained to ensure zero total force and torque
for each disc, consistent with these particles being swimmers driven
by their own internal dynamics and not subject to externally imposed
force monopoles or torques), as well as all higher modes of the
velocity (as prescribed by the tangential swimming function
$S$). Unknowns to be calculated are the lowest two modes of the
velocity, giving the solid body translational velocity $\vecv{V}_i$
and rotational speed $\Omega_i$.  (Also calculated as part of this
process are all higher modes of the force, but we do not report
these). The particle positions and swim directions are then updated
according to $\vecv{R}_i\to\vecv{R}_i+Dt\vecv{V}_i$ and
$\alpha_i\to\alpha_i + Dt\Omega_i$.

In this way we exactly (once the numerical mode numbers have been
converged upon) solve for the Stokes solution in the fluid outside all
the disks, subject to the boundary conditions of the prescribed
tangential velocities on the disk edges. Out of this calculation
emerge all effects of hydrodynamic interaction, including long range
power law propagators, short ranged lubrication effects, and the
physics on all lengthscales in between.

For active particles, the P\'eclet number $\Pe$ is defined as the time
taken for a particle thermally to diffuse a distance equal to its own
radius, divided by the time taken for it to swim the same distance.
Here we have assumed this to be infinite, as in
Refs.~\cite{tailleur2008,fily12,redner13a}, suppressing Brownian
motion entirely and considering only the deterministic hydrodynamics
defined above. This is the relevant limit physically for many active
suspensions of, {\it e.g.,} swimming bacteria. We do, however, return
to comment further on P\'eclet number below.

With less justification a priori, we also ignore any intrinsic
tumbling dynamics of the particles, such that within our model angular
reorientation occurs only via hydrodynamic interactions as the
particles scatter off each other. However, as we will explain further
in the results section below, we expect this neglect of tumbling to be
unimportant, because at the high volume fractions of interest here we
find that the relevant physics comes from the high reorientation rate
set by these scattering events, which would only be increased still
further by the explicit inclusion of tumbling events.

In the literature both spherical (3D)~\cite{Blake} and cylindrical
(2D)~\cite{Blake2D} versions of this squirmer model have been studied.
For computational efficiency we use the 2D model~\cite{Blake2D}, which
in fluid mechanical terms effectively corresponds to infinitely long
(in the $z$ direction) squirming cylinders propelling themselves in
the plane $x-y$ that is orthogonal to their length.  While this
geometry is of course to some extent artificial (as in all studies of
reduced dimensionality), it is important to realise that concerns
about the famous Stokes catastrophe of 2D hydrodynamics are not
relevant here. That catastrophe applies specifically to the case of a
particle subject to a force monopole, whereas the swimmers of interest
here are force free overall, experiencing only higher order force
multipoles. Indeed in a previous publication~\cite{fielding14} we
carefully compared this 2D case of infinite cylinders to a more
realistic 3D scenario of squirming disks of finite thickness ({\it
  i.e.} highly flattened cylinders) in a highly viscous film of the
same thickness, surrounded by a fluid of much lower viscosity.
Reassuringly we found no qualitative difference between these cases,
giving confidence that our 2D calculations provide a good
representation of the physics of suspensions of squirmers with
hydrodynamics.

\section{Units and parameter values}
\label{sec:parameters}

Table~\ref{parameters} summarizes the parameters of the problem,
defining the simulation geometry, particle size, modified
Lennard-Jones potential and the dynamics of the three models.
Throughout the paper we work in units of length in which the particle
radius $R=1$, of modulus in which the potential energy $\epsilon=1$,
and of time in which the drag coefficient $\gamma=1$ (for Brownian
particles) or the Newtonian viscosity $\eta=1$ (for hydrodynamic
squirmers).

Because of the heavy computational cost of the hydrodynamic
simulations (recall that we solve the Stokes equation fully in the
plane, recovering out of this calculation both far field power law
propagators, near field lubrication effects, and the physics on all
lengthscales between these), the number of squirmers that we can
feasibly simulate is strongly constrained. Here we choose $N=242$. The
more obvious choice of $N=256$ carries a greater risk of lattice
defects at area fractions approaching close packing. We have also
checked that the phenomena we report are robust against doubling
$N$. For the Brownian dynamics (both passive and active) many more
particles would be possible because of the much lower computational
demand of those simulations: up to $N=10^5$ or $N=10^6$ Brownian
particles can be simulated in run times comparable to those for $N=242
$ hydrodynamic squirmers.  However we restrict ourselves to $N=242$
for the PBP and ABP as well, in order to ensure a direct comparison
between all the models.  We return at the start of
Sec.~\ref{sec:results} below to comment carefully on the extent to
which one might expect to capture a system's phase behaviour with such
a small number of particles.

The lengthscales $\sigma_{\rm LJ}$ and $\sigma_{\rm h}$ associated
with the potential are set as discussed in Sec.~\ref{sec:potential}
above and summarised again in table~\ref{parameters}.

An important parameter that then remains to be varied in the
simulations for all three models is the area fraction $\phi$.  Because
of the strong repulsive force characterized by the small shell length
$\sigma_{\rm h}$, it is reasonable to work with the effective area
fraction defined above rather than the bare area fraction $\phi_r=N\pi
R^2/L^2$.  The real area fraction in practice of course depend on the
ratio between the temperature (or active velocity) and the potential
energy $\epsilon$.  However the steepness of the repulsive shell's
potential renders this effect small in practice. Therefore in what
follows, unless stated explicitly otherwise, we report the effective
area fraction.  Our simulation method allows us to study high area
fractions even for hydrodynamic squirmers, and accordingly we shall
explore right across the range from dilute towards close packed.

Besides the area fraction, the other important parameters that remain
to be varied are as summarised follows. For the passive Brownian
particles we have the adimensionalised translational diffusional
coefficient $\tilde{D}$, which is also equivalent to the
adimensionalised thermodynamic temperature.  For the active Brownian
particles we have the adimensionalised active swim speed $\vat$ and
adimensionalised rotational diffusion coefficient
$\zeta^{-1}$. Finally for the hydrodynamic squirmers we have the
adimensionalised active swim speed $\vat$ and the stresslet parameter
$\beta$. We shall return to discuss in more detail the physical
meaning of each of these adimensional quantities in each of the
individual subsections concerning PBP, ABP and HS below.

\section{Initial conditions}
\label{sec:initial}

In each run the system is initialised by placing the particles at
random positions and with orientations likewise chosen at random from
a top hat distribution between $0$ and $2\pi$.  Obviously some
particles will inevitably overlap after this procedure.  Overlaps are
then removed by evolving the system with zero-temperature dynamics in
a landscape of mutual particle repulsion. 

We anticipate (but have not checked explicitly) that in any regions of
the phase diagram where our simulations attain steady state, that
state would be independent of the initial conditions. However, as we
shall discuss below, at low temperatures and area fractions we
typically find a regime of coarsening dynamics in which the average
cluster size slowly increases with time. Due to the very slow dynamics
we have not been able to attain steady state in any feasible run time
in that regime, and these states must inevitably depend on the initial
conditions.

\section{Measured quantities}
\label{sec:measures}

In this section we describe the various quantities that we measure in
our simulations in order to analyse clustering and phase behaviour
across the different models.

\subsection{Snapshots}

In each case we first make a simple visual observation of snapshots of
the system's configuration, on a grid of values of the area fraction
and adimensionalised temperature (or activity parameter).  Unless
otherwise stated the snapshots shown are in steady state, verified by
confirming that the time signals of the mean cluster size, mean
particle velocity, {\it etc.} have attained their final values (which
are of course subject to fluctuations). Where stated, in some cases
of extremely slow coarsening at low temperature, the snapshots shown
are for the longest times that are feasibly accessible numerically.

\subsection{Cluster analysis}
\label{sectionclusters} 

We statistically analyse the degree of clustering in the system using
a single-link algorithm, the details of which are described in
Refs.~\cite{Jainclustering,johnson67}. Any two particles are taken to
be in the same cluster if their separation $r$ is smaller than a
cutoff value $r=r_c$.  Indeed it is worth reiterating that any two
groups of particles are taken to belong to the same cluster even if
only a {\em single} pair of particles, one in each group, satisfies
this cutoff condition. We choose a value for the cutoff separation
$r_c=2.4R$, which just exceeds the value of the particle separation at
the minimum in the potential at $r=2^{1/6} 2 R\approx 2.24 R$. We
have checked the robustness of our results to reasonable variations in
the value of this cutoff. Common practice in the
literature~\cite{PhysRevLett.111.245702, prymidis15} is to use a
cutoff smaller than the first minimum of the radial distribution
function, and we have followed that convention here.

Using this algorithm we measure the cluster size distribution $p(n,t)$
at any instant of time $t$, and the first moment $\bar{n}(t)$ of this
distribution, which gives the mean number of particles per cluster. We
usually report this quantity once the system has reached a
statistically steady state after many time units, and time-averaged
over a time interval $\Delta t$ to ensure good statistics. Typically
$\Delta t=3 10^3 R^2/3D$ for PBP, $\Delta t =3\times 10^3 R/3v_a$ for
ABP and $\Delta t= 10^3 R/3v_a$ for HS.  In a few, explicitly
stated, cases of extremely slow coarsening at low temperature the
cluster size is reported at the longest times that are feasibly
accessible numerically, again time averaged over $\Delta t$.  

The instantaneous (non time averaged) behaviour of $\bar{n}(t)$ as
function of time $t$ (after initialising the system in a disordered
configuration at time $t=0$ as described in Sec.~\ref{sec:initial})
will also be used to to characterise the dynamical coarsening of
clusters in the system.

\subsection{Particle number fluctuations}

In order to gain insight into fluctuations in the local area fraction
of particles, we divide the simulation box into a grid of $G\times G$
cells each of dimension $L_G=L/G$, with a typical value of $G=5$. We
then define the instantaneous local particle number in each cell $i$
of the grid as $N_i(t)$ and write this as the sum of the average value
plus a fluctuation away from this:
\be
N_i(t)=\frac{N}{G^2}+\delta N_i(t),
\ee
and measure the number fluctuations as
\be
\delta_N=\langle\sqrt{\langle\delta N_i(t)^2\rangle_i}\rangle_t,
\ee
which we typically report scaled by $\sqrt{\bar{N}}$, with
$\bar{N}=\langle N_i(t)\rangle_i=N/G^2$. Here $\langle \cdot
\rangle_t$ represents an average over time, and $\langle \cdot
\rangle_i$ an average over particles.

Alternatively we can think in terms of the local area fraction
$\phi_i(t)= N_i(t)\pi R^2/L_G^2=N_i(t) \phi G^2/N$, which for a given
grid division $G$ and total number of particles $N$ is related to the
local number of particles $N_i(t)$ by a constant factor that depends
only on the global average area fraction $\phi$. 
The local area fraction fluctuations are then defined as
\be
 \delta_\phi=\langle\sqrt{\langle\delta \phi_i(t)^2\rangle_i}\rangle_t.
\ee
To connect these two measures we note simply that $\delta_\phi/\phi=
G^2\delta_N/N$. In what follows below we sometimes find it convenient
to report $\delta_N$, which more effectively shows fluctuations at low
$\phi$, and sometimes $\delta_\phi$, when focusing on higher $\phi$.
(The values of $G=5$ and $N=242$ are the same in all our simulations.)

\subsection{Hexatic order parameter}

While the cluster size distribution and particle number fluctuation
measures described so far are effective at discerning the degree of
aggregation, they are poorly suited to differentiating between liquid
and solid phases. We therefore define a hexatic order parameter
$\Psi_6$ to do this.

We start by defining an instantaneous particle-based local hexatic
order, for a given particle $j$ at any time $t$:
\begin{eqnarray}
\Psi_{6j}(t)=\frac{1}{\hat{n}_j}\sum_{k=1}^{\hat{n}_j}{e^{i 6 \theta_{jk}(t)}}.
\end{eqnarray}
Here $\hat{n}_j$ is the number of neighbours of particle $j$, defined
as those particles inside a threshold separation $r<r_c=2.4R$, while
$\theta_{jk}(t)$ is the angle of the bond between particles $j$ and
$k$.

From this instantaneous particle-based measure of local hexatic order
one can then define various different measures of global hexatic
order.  See~\cite{ziren10} for two definitions. Here we focus
specifically on the one that averages $\Psi_{6j}$ in time $\langle
\cdot \rangle_t$ first for each particle, and then averages the
modulus of this quantity over all particles $\langle \cdot \rangle_j$
to give
\begin{eqnarray}\label{psi_def}
\Psi_6&=& \langle |\langle \Psi_{6j}(t) \rangle_t |\rangle_j.
\end{eqnarray} 
For a single perfectly hexatically ordered crystal we expect
$\Psi_6=1$, while for a totally disordered phase $\Psi_6=0$.

\section{Results}
\label{sec:results}

We now present the results of our simulations.  Even though the
equilibrium phase behaviour of passive Lennard-Jones (LJ) particles
has been mapped out previously, both in 2D~\cite{Binder1990} and
3D~\cite{Hansen69}, we shall nonetheless start in
Sec.~\ref{sec:PBPresults} by studying the equilibrium phase behaviour
with our modified LJ potential, which includes the repulsive shell
added to avoid particle overlaps for the squirmers.  We do so in order
to allow in later sections~\ref{sec:ABPresults}
and~\ref{sec:HSresults} a truly direct comparison of the
non-equilibrium phase behaviour of active particles - both without and
with hydrodynamics - with the equilibrium phase behaviour of their
passive counterpart experiencing exactly the same interaction
potential, including this repulsive shell.  (We do not perform
simulations of passive particles with hydrodynamics, because the phase
behaviour of equilibrium systems depends only on the static
interaction potential and not the dynamics.)

Because the number of particles that we can access for the
hydrodynamic squirmers is heavily limited by computational cost (we
use $N=242$), it is furthermore important to establish in this more
familiar equilibrium context to what extent phase behaviour can
feasibly be discerned with such a small system size.  Although this
number of particles is of course very small for systems without
hydrodynamics, by the standards of today's computational capabilities,
it is worth noting that the early papers to map out the phase
behaviour of Lennard-Jones and hard sphere systems did so using
typically $200-1000$
particles~\cite{Binder1990,Hansen69,Alder1960}, even
in 3D. Indeed these early works established that surprisingly small
numbers of particles are needed to map the overall features of the
phase diagram correctly. Accordingly we claim that our $N=242$
simulations can capture the main features of phase behaviour.
However we do not claim full quantitative statistical analysis of two
phase regions, or even less of critical points, to be a realistic goal
in this work.

\subsection{Passive Brownian particles}
\label{sec:PBPresults}

\begin{figure}[h] 
\centering
\includegraphics[width=8.5cm]{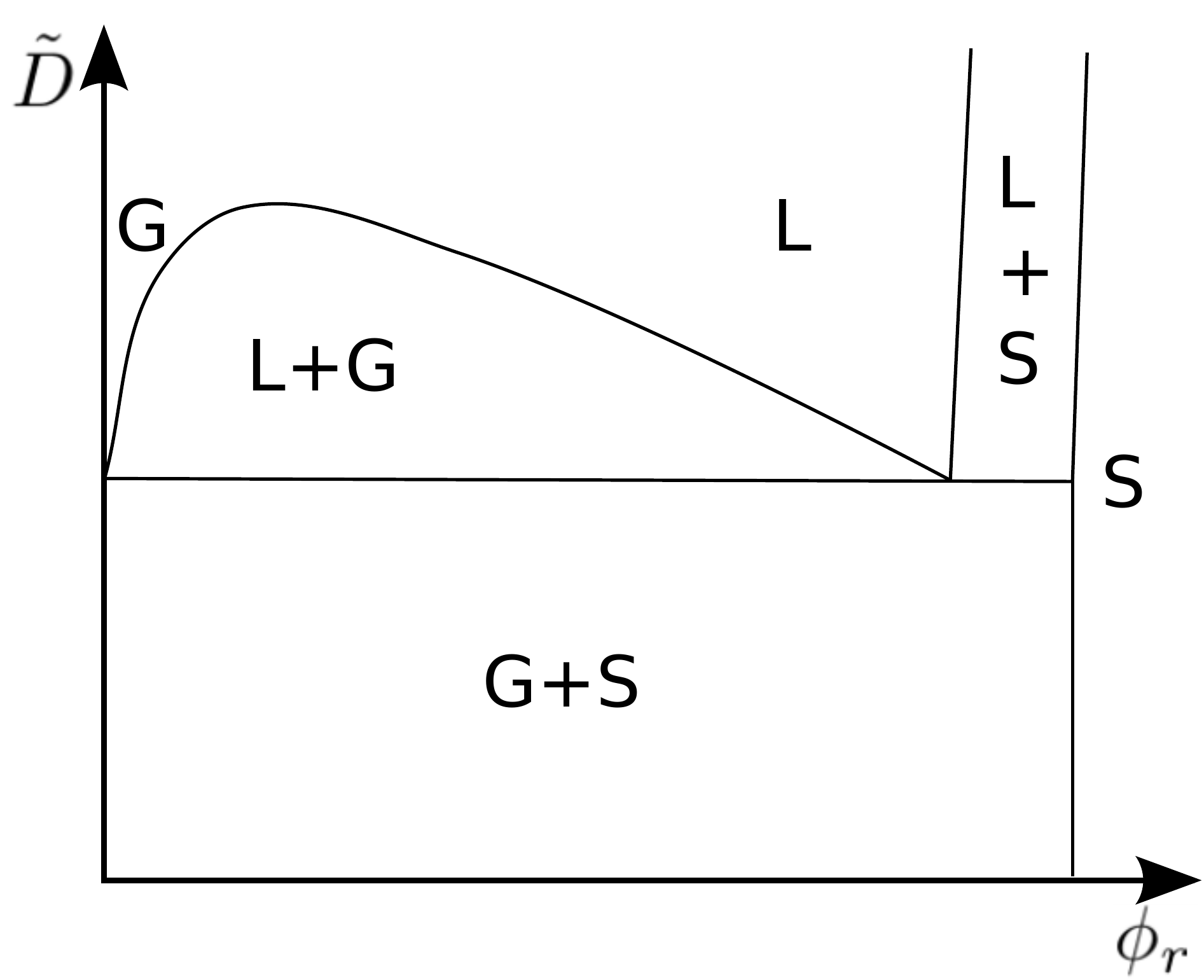}
\caption{\label{phasediagsketch} Sketch of the main qualitative
  features of the equilibrium phase diagram of disks with a
  Lennard-Jones interaction potential $V_{\rm
    LJ}$~\cite{Hansen69,tang02}.  S=solid, L=liquid, G=gas.}
\end{figure}
\begin{figure*}
\centering
\includegraphics[width=14cm]{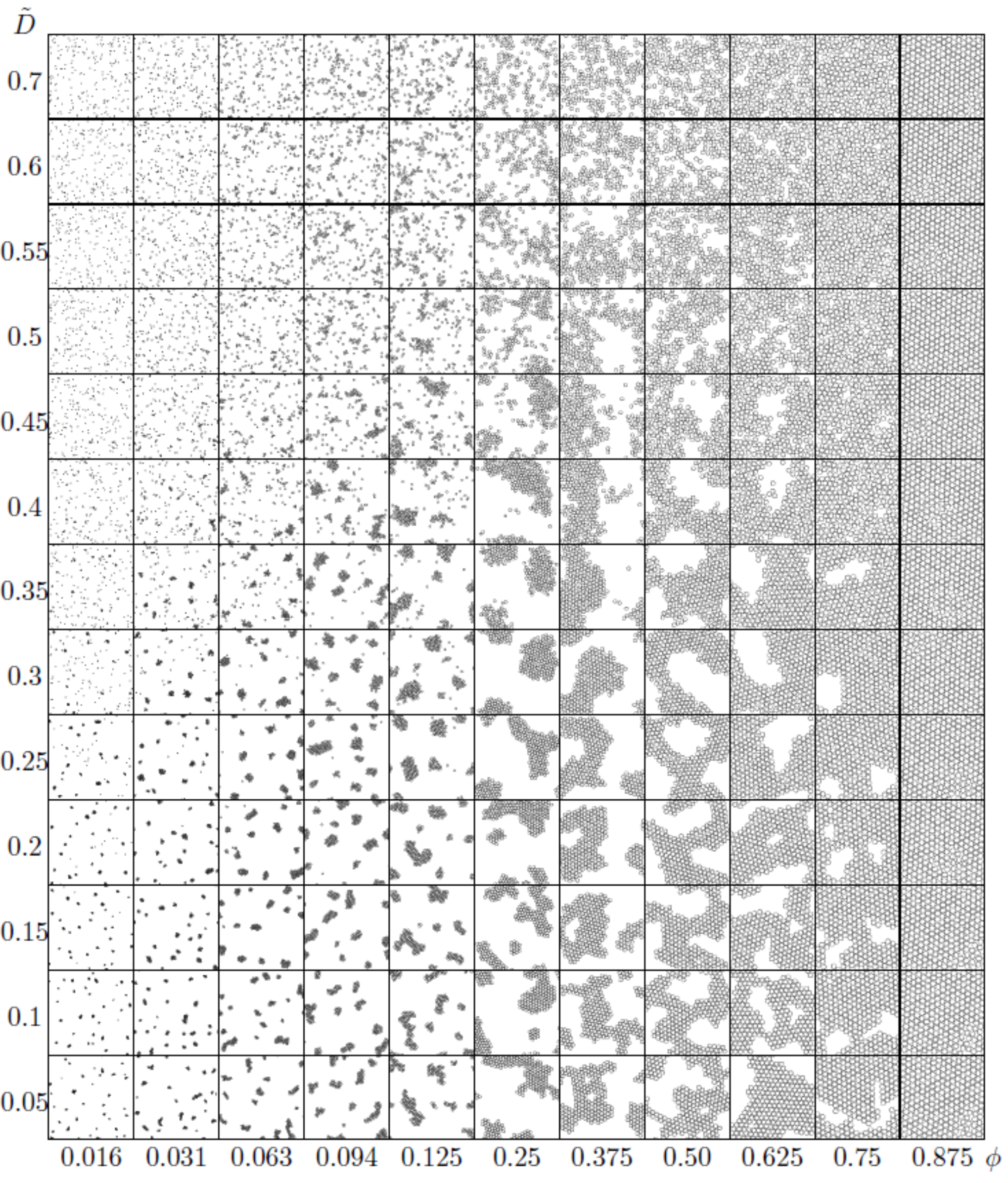}
\caption{\label{passiveconfigmap} Passive Brownian particles.
  Snapshots of the system's configuration on a grid of values of the
  scaled thermal diffusion coefficient $\tilde{D}$ and area fraction
  $\phi$. Each snapshot is taken at a long time $t=10^3 R^2/D$ after
  the system was initialised at $t=0$ in a random state as described
  in Sec.~\ref{sec:initial}.  Note that the scale is nonlinear at the
  largest $\tilde{D}$ and at low $\phi$. }
\end{figure*}

\begin{figure}[h!] 
\centering
\subfigure[(a)]{\includegraphics[width=8.0cm]{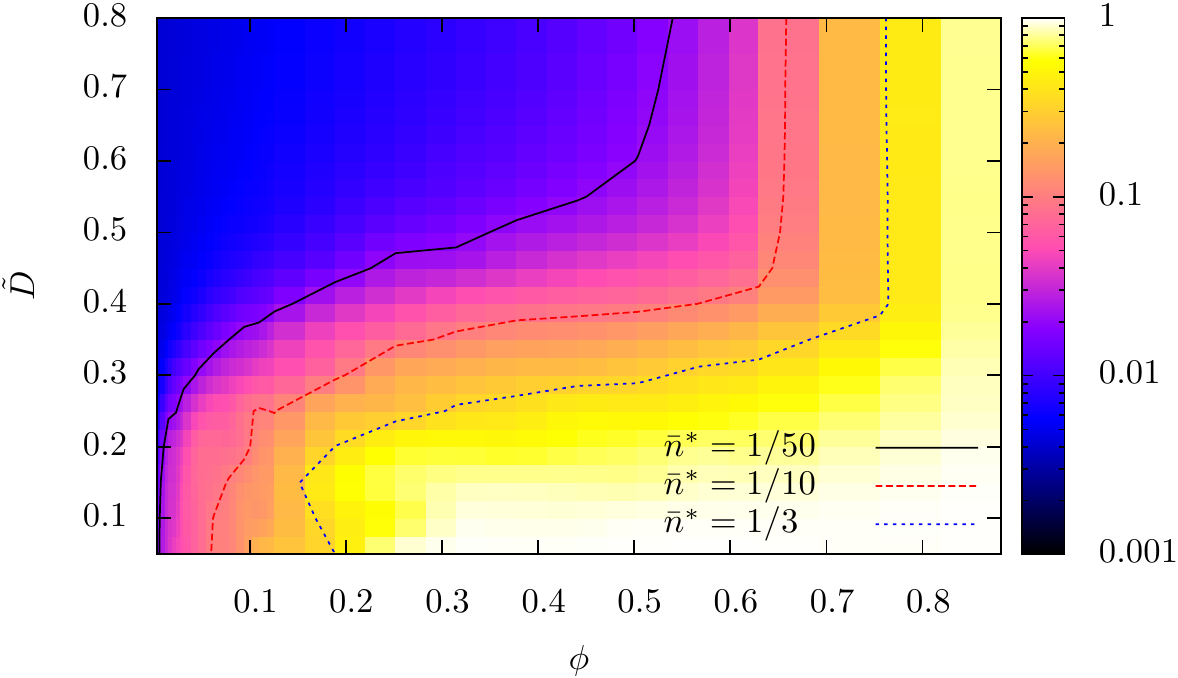}}
\subfigure[(b)]{\includegraphics[width=8.0cm]{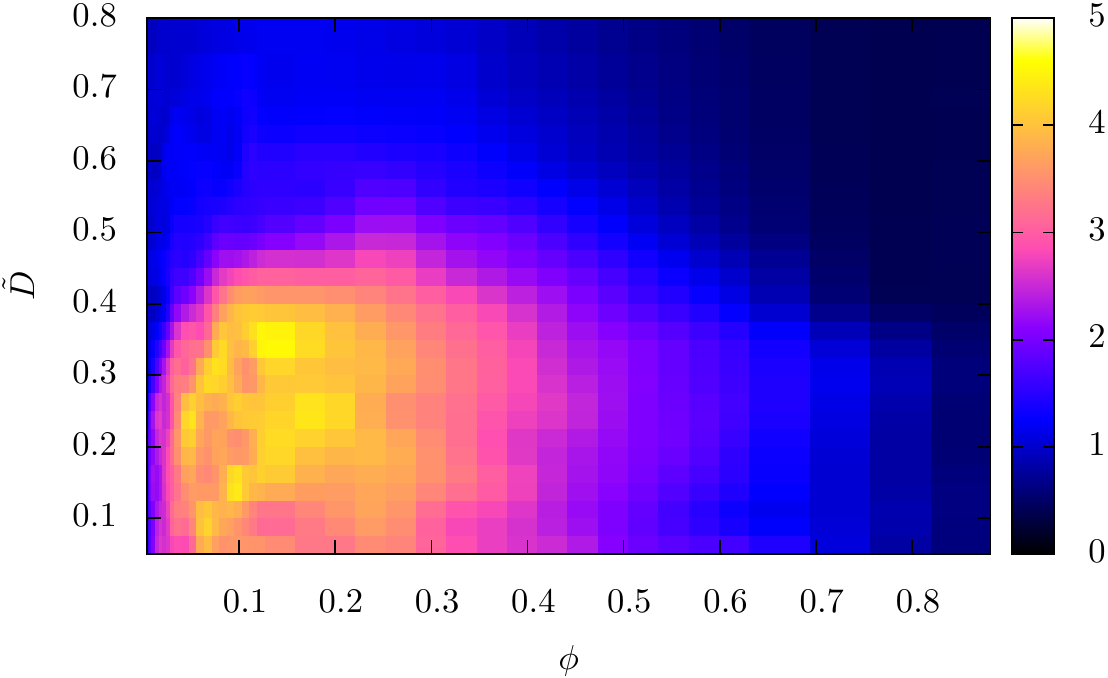}}
\subfigure[(c)]{\includegraphics[width=8.0cm]{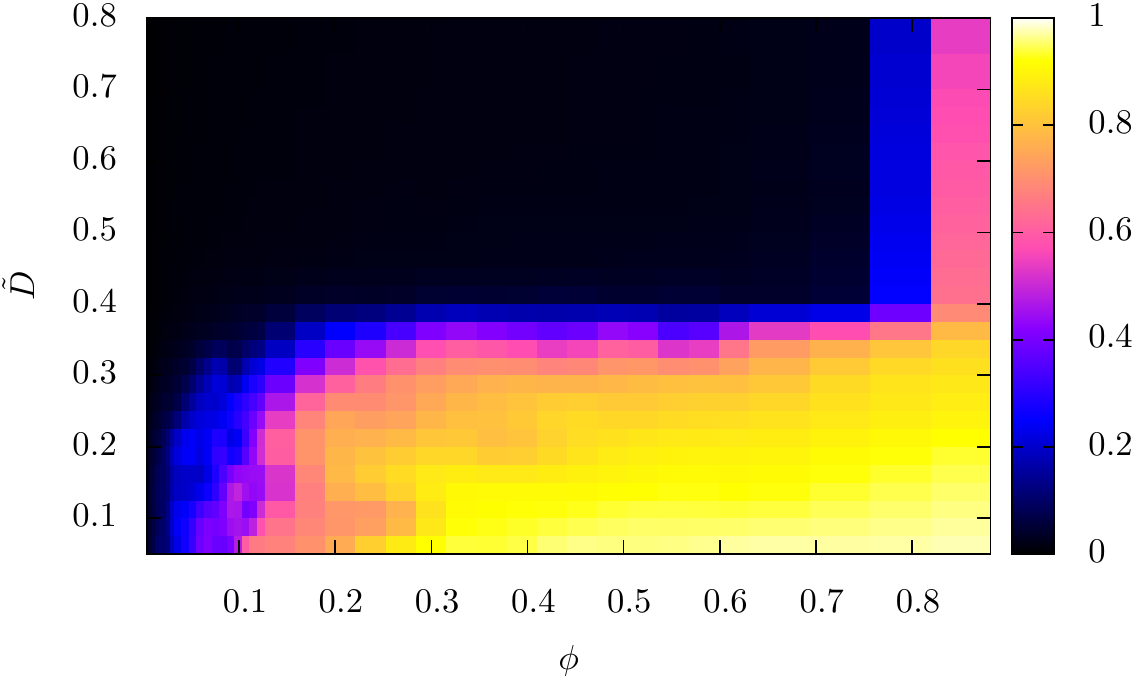}}
\caption{Phase diagram of passive Brownian particles mapped using (a)
  normalised mean number of particles per cluster
  $\bar{n^*}=\bar{n}/N$, (b) particle number fluctuations
  $\delta_{N}/\sqrt{\bar{N}}$ for $G=5$ and c)
  hexatic order parameter $\Psi_6$. \label{fig:PBPphasediag}}
\end{figure}

\begin{figure}[h]
\centering
\includegraphics[width=7.5cm]{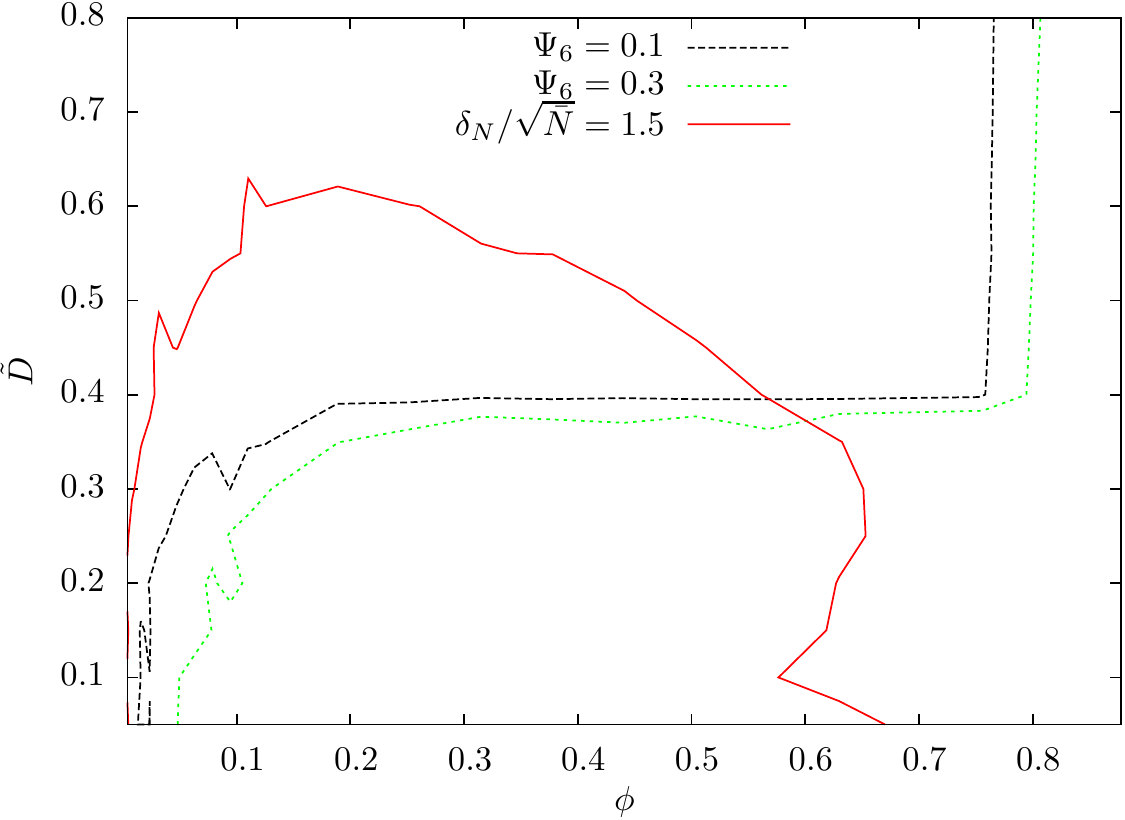}
\caption{Representative isolines of particle number fluctuations $\delta_N/\sqrt{\bar{N}}=1.5$ and $\Psi_6=0.1,0.3$. Regions broadly consistent with the sketch of Fig.\ref{phasediagsketch} are evident.
  \label{contourspsidn} }
\end{figure}

As an aid to the reader's memory we start by sketching in
Fig.~\ref{phasediagsketch} the main features of the equilibrium phase
diagram of particles experiencing unmodified LJ interactions, in the
plane of adimensionalisation diffusion coefficient $\tilde{D}$ versus
the area (or volume) fraction $\phi$. Recall that
$\tilde{D}=D\gamma/\epsilon=k_{\rm B}T/\epsilon$ measures the thermal
energy of the system relative to the characteristic energy of the
modified Lennard-Jones potential. As noted above this phase diagram
has been comprehensively mapped out
previously~\cite{Binder1990,Hansen69}.  Indicated are the single
phase regions of gas (G), liquid (L) and solid (S), and the phase
coexistence regions G-L, G-S, L-S.  We emphasize that
Fig.~\ref{phasediagsketch} is a rough qualitative sketch only, with no
attempt at quantitative accuracy.

Turning now to our modified Lennard-Jones potential, we show in
Fig.~\ref{passiveconfigmap} snapshots of the system's configuration on
a grid of values $\tilde{D}$ and area fraction $\phi$.  All snapshots
shown are for the longest time that is feasibly accessible
computationally. At low temperatures,$\tilde{D}\lae 0.4$, aggregated
domains are evident across the full range of area fractions.  These
coarsen very slowly as a function of time, and at the lowest area
fractions and temperatures (typically in the rectangle of phase space
with $\phi < 0.125$ and $\tilde{D}<0.4$) this process cannot be
evolved fully to steady state within a feasible run time. (For
example, an order of magnitude estimate suggests that for the
simulation at $\phi=0.0125$ and $\tilde{D}=0.05$ to have fully
coarsened, a run time of about 6 months involving $O(10^9)$ timesteps
would have been needed.) The snapshots in this region are therefore
not in true steady state, though we are confident that the qualitative
features will not change even for longer run times, which would simply
result in slightly larger clusters. At higher temperature we have
checked that the system is in steady state.

Simple visual inspection of these snapshots suggests an overall
structure to the phase diagram that indeed captures the main features
of the sketch in Fig.~\ref{phasediagsketch}. In particular, a G-S
coexistence region is visually apparent for $\tilde{D}$ less then
about $0.35$. Above this, a G-L coexistence region extends up towards
what resembles a critical point somewhere in the region of $\phi=0.4$
and $\tilde{D}=0.55$. However we do not focus on this critical point
in detail because we do not have sufficient particles in our
simulation to characterise it properly. For high area fractions,
roughly $\phi>0.7$, solid-like ordering is evident even at high
temperatures.  This area fraction for the onset of solid ordering is
lower than that expected for particles with conventional LJ
interactions, but is consistent with the fact that our modified LJ
potential includes an effectively hard shell repulsion at $r=2R$.

To try map the phase diagram more quantitatively, we show in
Fig.~\ref{fig:PBPphasediag} colour maps of the statistical measures
introduced in Sec.~\ref{sec:measures} above. Subfigure (a) shows the
mean number of particles per cluster, normalized by the total number
of particles in the system: $\bar{n}^*\equiv \bar{n}/N$.  Note that a
system-spanning homogeneous phase with all particles in same cluster
would have $\bar{n}^*=1$, while a gas phase with no clusters would
have $\bar{n}^*=1/N$, which is small for large $N$. With this in mind
we have also marked on the colourmap isolines of
$\bar{n}^*=1/50,1/10,1/3$, which correspond to
$\bar{n}=4.84,24.2,80.7$ for our $N=242$. We suggest that the region
in between the first two of these corresponds to one of clusters
 of a finite size $\bar{n}>1$, but which do not yet span the
full system size $\bar{n}^*=1$. Comparison with simulations of larger
$N$ would however be needed to substantiate this claim properly. Note
that, as above for the snapshots, in the region roughly given by
$\phi<0.25$ and $\tilde{D}<0.4$ the clusters are still coarsening so
that waiting for an even longer run-time would result in a larger
$\bar{n}$.

Subfigure (b) shows the particle number fluctuations
$\delta_N/\sqrt{\bar{N}}$, measured by dividing the simulation box
into a grid of $G\times G=5\times 5$ subunits. Clearly apparent is a
region with strongly enhanced number fluctuations, which we suggest
can be interpreted as a region of phase coexistence.  However this
measure provides no means of distinguishing between a phase
coexistence that involves only disordered phases, $G-L$, from one
involving the ordered solid phase, $G-S$ or $L-S$. In any case it is
strongly suppressed at high volume fractions $\phi>0.7$ and so fails
altogether as an indicator of phase separation in this regime.

To try identify these solid regimes, we show in subfigure (c) a colour
map of the hexatic order parameter. The bright region for $\phi>0.75$
clearly indicates the presence of crystalline ordering. We further
suggest the region $\tilde{D}<0.35$ to be one of $G-S$ coexistence (or
fully solid at high $\phi$, but this cannot be discerned separately),
consistent with visual inspection of the snapshots in
Fig.~\ref{passiveconfigmap}.  (For small area fractions the solid
component is too weak to show in this representation. As noted above,
however, in this part of the phase diagram coarsening is so slow that
this system does not attain a true steady state within the run time.)
We also suggest that for $\tilde{D}>0.35$ the region $0.75<\phi<0.8$
is one of $L-S$ coexistence, and that of $\phi>0.8$ a homogeneous $S$
phase. We emphasize however that these boundaries are suggestive only,
and that simulations with far more particles would be needed to
provide quantitatively conclusive evidence.

Finally we assemble in Fig.~\ref{contourspsidn} a representative
isoline of the particle number fluctuations $\delta_N/\sqrt{\bar{N}}=1.5$
together with two representative isolines of the hexatic order
parameter $\Psi_6=0.1,0.3$, revealing a phase
diagram that indeed qualitatively resembles the sketch in
Fig.~\ref{phasediagsketch}.

Although our main aim in this section has been to report equilibrium
phase behaviour, we return finally to reiterate that in the rectangle
of phase space with $\phi < 0.25$ and $\tilde{D}<0.4$ the system
doesn't equilibrate in any feasible run time but rather shows a slow
coarsening behaviour in which the typical cluster size increases as a
power law $\bar{n}\sim t^\beta$. A value for the exponent $\beta=2/3$
provides an excellent fit to the data (not shown), consistent with the
expectation~\cite{chaikin95} of a linear domain size growing as $L\sim
t^{1/3}$.  Although we have been careful only to extract data for the
coarsening exponent in regimes where the typical cluster size is less
than the system size, a more detailed study of the model's coarsening
dynamics, for which finite size effects should be investigated
carefully, is out of the scope of this paper.

\subsection{Active Brownian Particles}
\label{sec:ABPresults} 

In the previous subsection we mapped out the equilibrium phase
behaviour of passive Brownian particles (PBPs) in the plane of area
fraction $\phi$ and adimensional diffusion coefficient
$\tilde{D}=D\gamma/\epsilon=k_{\rm B}T/\epsilon$, which measures the
system's thermal energy compared with the characteristic energy
$\epsilon$ of the modified Lennard-Jones potential. As discussed, we
recovered the main features of phase behaviour expected from previous
studies of Lennard-Jones particles, with some differences consistent
with our use of an additional steep repulsive contribution $V_{\rm
  h}$.

We turn now to active Brownian particles (ABPs). Although the phase
behaviour of ABPs with unmodified Lennard-Jones interactions has been
studied previously~\cite{redner13b}, we will map it out here for our
modified Lennard-Jones potential using the same small number of
particles $N=242$ as will be feasible to explore computationally for
the hydrodynamic squirmers in Sec.~\ref{sec:HSresults} below. This
will then enable a truly direct comparison between ABPs, which lack
hydrodynamic interactions, and squirmers, which have them.

As discussed above, the ABPs move via ballistic swimming at speed
$\vat$ combined with stochastic reorientational dynamics with angular
diffusion coefficient $\Dr$. The latter is intended as a continuous
time model of (for example) discrete tumbling events, in some species
of bacteria, or reorientations due to hydrodynamic scattering off
other swimmers. It is not intended to model true thermal angular
diffusion, which would be much smaller in comparison. Indeed, as noted
above, we consider here the athermal limit in which any thermal
contribution to $\Dr$ is zero, which is the relevant limit for many
swimming microbes. 

Despite this lack of true thermal translational diffusion, an obvious
question to ask is whether the activity of the ABPs (ballistic
swimming combined with stochastic reorientation) plays a role somewhat
akin to the thermal diffusive dynamics of the PBPs, such that the
non-equilibrium phase behaviour of the ABPs is in some degree
analogous to the equilibrium phase behaviour of the PBPs. For example,
we might intuitively anticipate phase separation at small $\va$ due to
the attractive effects of the Lennard-Jones potential, with a return
to homogeneous behaviour at large $\va$ where the activity is strong
enough to overcome that attraction.  In what follows, therefore, we
will seek to compare the equilibrium phase behaviour of PBPs in the
plane of $(\Dt,\phi)$, as mapped out above, with the non-equilibrium
phase behaviour of the ABPs in the plane of $(\va,\phi)$, with $\va$
first suitably adimensionalised.

However, whereas for PBPs the only two relevant parameters were indeed
$\Dt$ and $\phi$, for ABPs the rotational diffusion coefficient $\Dr$
is also important, alongside $\va$ and $\phi$.  The reason for this is
as follows. In the case of spherical PBPs rotational diffusion, though
surely present physically, is irrelevant: each particle is
rotationally symmetric and the translational dynamics is accordingly
independent of particle orientation.  In contrast, for ABPs the
translational dynamics depends strongly on the orientational dynamics,
because the instantaneous preferred swimming direction of any particle
is (by definition) prescribed by its instantaneous orientation. In
view of this we must actually consider the phase behaviour of ABPs in
the plane of $(\va,\phi)$ for several different values of $\Dr$, with
both $\va$ and $\Dr$ first suitably adimensionalised.

To adimensionalise the rotational diffusion coefficient $\Dr$ we
define a parameter~\cite{fielding14} 
\be
\zeta=\frac{\va}{\Dr Rf(\phi)}= \frac{\tauo}{\tauc(\phi)},
\ee
which is the ratio of the characteristic decorrelation time of the
particle's orientation (swim direction)
\be
\tauo=\frac{1}{\Dr}
\ee
with the characteristic time interval between particle collisions
\be
\tauc=\frac{R f(\phi)}{\va}.
\ee

\begin{figure*}
\centering
\includegraphics[width=14.0cm]{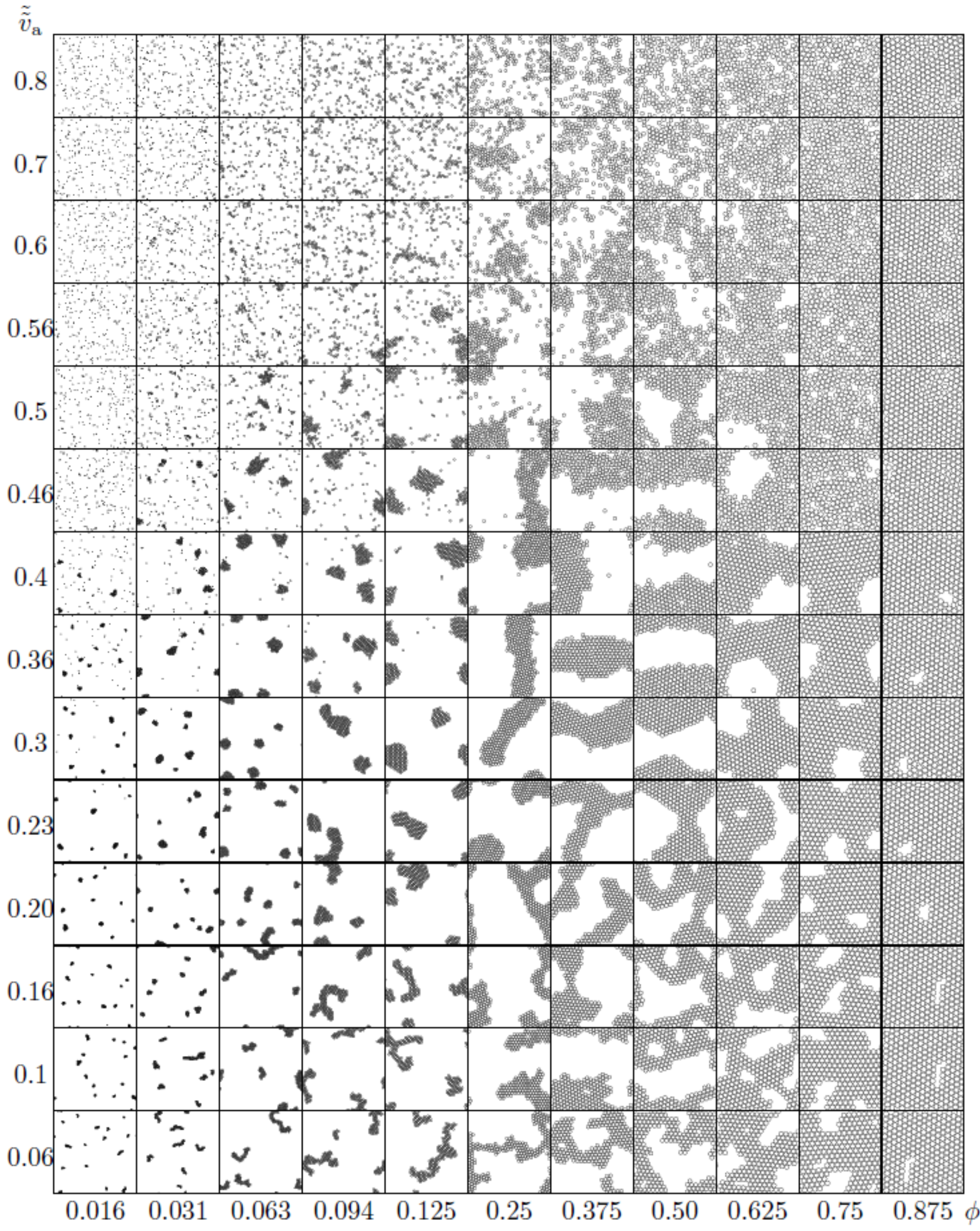}
\caption{\label{snaps_ab_z_.2} Active Brownian particles, $\zeta=0.2$.
  Snapshots of the system's configuration on a grid of values of the
  scaled active swim speed $\va$ and effective area fraction $\phi$.
  Each snapshot is taken at a long time $3\times 10^3 v_a/R$ after the
  system was initialised at $t=0$ in a random state as described in
  Sec.~\ref{sec:initial}.  Note that the scale is nonlinear at the
  largest $\vatt$ and at low $\phi$.}
\end{figure*}

\begin{figure*}
\centering
\includegraphics[width=14.0cm]{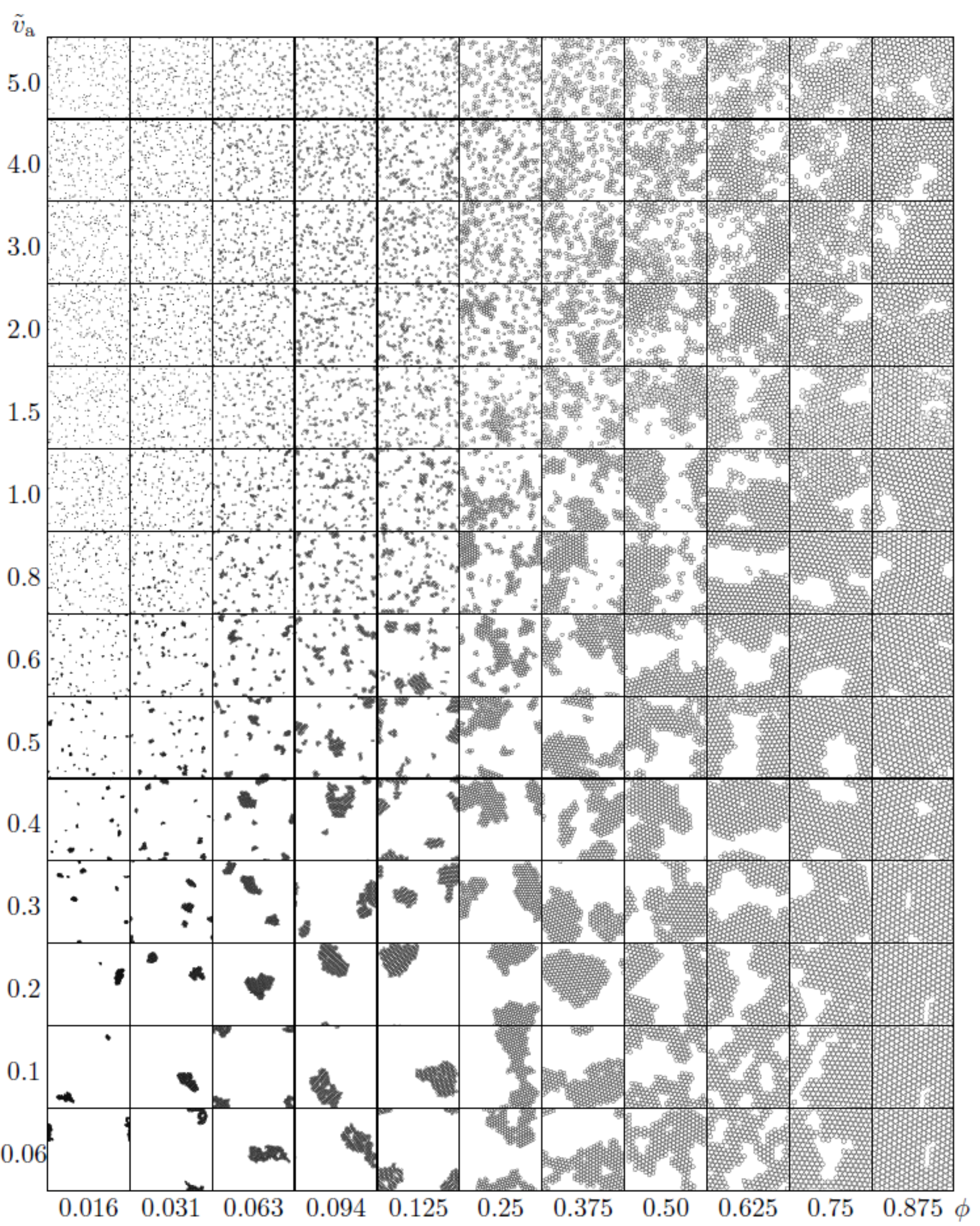}
\caption{\label{snaps_ab_z_100}Active Brownian particles,
  $\zeta=200.0$.  Snapshots of the system's configuration on a grid of
  values of the scaled active swim speed $\va$ and effective area
  fraction $\phi$. Each snapshot is taken at a long time $3\times 10^3
  v_a/R$ after the system was initialised at $t=0$ in a random state
  as described in Sec.~\ref{sec:initial}.  Note that the scale is
  nonlinear at the largest $\vat$ and at low $\phi$.}
\end{figure*}

Here $f(\phi)$ is an adimensional function that we expect to decrease
with increasing $\phi$. In the dilute gas limit of small $\phi$ we would estimate $f=\pi/\phi$
from mean free path arguments, while for more crowded systems with
$\phi=O(1)$ we might instead simply assume $f=O(1)$.  Because our main
focus in this paper concerns moderately to highly crowded systems,
$\phi=O(1)$, for simplicity we set $f=1$ throughout.  Accordingly we
work simply with
\be
\zeta=\frac{\va}{\Dr R}.
\ee

In the regime $\zeta\ll 1$, particles change their orientation many
times during a typical interval between collisions.  Therefore the
dynamics during an interval $0<t<\tauc$ between any two collisions
will comprise only a proportionately much shorter regime of ballistic
motion at speed $\va$ for times $0<t<\tauo\ll \tauc$, followed by a
longer diffusive regime for times $\tauo<t<\tauc$. In this way the
``microscopic'' (pre-collision) dynamics that feed forward to inform
the collisional dynamics are predominantly diffusive, with an
effective diffusion coefficient $\va^2/\Dr$. It might then be
reasonable to expect that the phase diagram of the ABPs will mirror
that of the PBPs with the scaled diffusion coefficient $\tilde{D}=D
\gamma/\epsilon$ of the passive system replaced its scaled counterpart
$\va^2\gamma/\Dr\epsilon$ for the active one.

In contrast, in the regime $\zeta\gg 1$ the dynamics of the ABPs will
be dominated by ballistic swimming over the entire time between
collisions, with the slow angular reorientation being relatively
unimportant in comparison. In this case it is less clear, upfront,
what correspondence might be expected between the phase diagram of the
ABPs with its equilibrium counterpart for the PBPs. As elaborated
further below, however, we will indeed find some correspondence in the
regime of low activity.

This existence of two different regimes of the dimensionless inverse
angular diffusion coefficient $\zeta$ points to two different possible
choices for adimensionalising the swim speed $\va$.  The first, which
we expect to be the natural choice in the regime $\zeta\ll 1$, takes
into account both ballistic swimming and angular reorientation by
recognising that $\va^2/\Dr$ has the dimensions of a translational
diffusion coefficient then writing $\vatt=\va^2\gamma/\Dr\epsilon$, by
direct analogy with $\tilde{D}=D\gamma/\epsilon$ for the PBPs.  

The second choice, which we expect to be the natural one in the regime
$\zeta\gg 1$, considers only the ballistic component of the swimming
dynamics, taking the ratio of $\va$ with the characteristic speed
$\epsilon/R\gamma$ that a particle would exhibit when subject to a
force of magnitude $\epsilon/R$ in a solvent of drag coefficient
$\gamma$. This gives an adimensionalised swim speed $\vat=\va
R\gamma/\epsilon$ that ignores the angularly diffusive component of
the dynamics as characterised by $\Dr$.  It is nonetheless still seen
as a possible counterpart of the $\tilde{D}$ of PBPs by writing
$\tilde{D}=D\gamma/\epsilon=(D/R)R\gamma/\epsilon$, and noting that
$D/R$ has dimensions of speed.  We note that for $\zeta=1$,
$\vat=\vatt$ and the two methods of adimensionalising $\va$ coincide,
as required.
  
With this preamble in mind, we now present results for the phase
behaviour of ABPs in the plane of $(\va,\phi)$ for two values of the
adimensional inverse diffusion coefficient $\zeta$, one in each regime
just described, with the swim speed $\va$ in each case first
adimensionalised in a manner suited to the value of $\zeta$ according
to the arguments just outlined. Where useful, we will also further
present results in the plane of $(\vat,\zeta)$ for different values of
$\phi$.

We start in Fig.~\ref{snaps_ab_z_.2} by showing long-time snapshots of
the system's configuration on a grid of values of $(\vatt,\phi)$ for a
low value of the scaled inverse diffusion coefficient $\zeta=0.2$.
Immediately apparent is that this non-equilibrium phase diagram
displays striking similarities to its equilibrium counterpart for the
PBPs in the plane of $(\tilde{D},\phi)$: recall
Fig.~\ref{passiveconfigmap}. In particular, a G-S coexistence region
is seen for values of $\vatt$ less then about $0.4$. Above this a G-L
coexistence region extends up towards what resembles a critical point
somewhere in the region of $\phi=0.4$ and $\vatt=0.7$, which is quite
close to that identified above for the PBP. For high area fraction,
solid-like ordering is also again evident. For this small value of
$\zeta$, then, the main features of the equilibrium phase behaviour of
PBPs transcribe to the non-equilibrium ABPs, with the scaled
temperature $\tilde{D}=\kB T/\epsilon$ replaced by the scaled swim
speed $\vat$.

\begin{figure}[h]
\centering
\subfigure[$\zeta=0.2$]{\includegraphics[width=8.5cm]{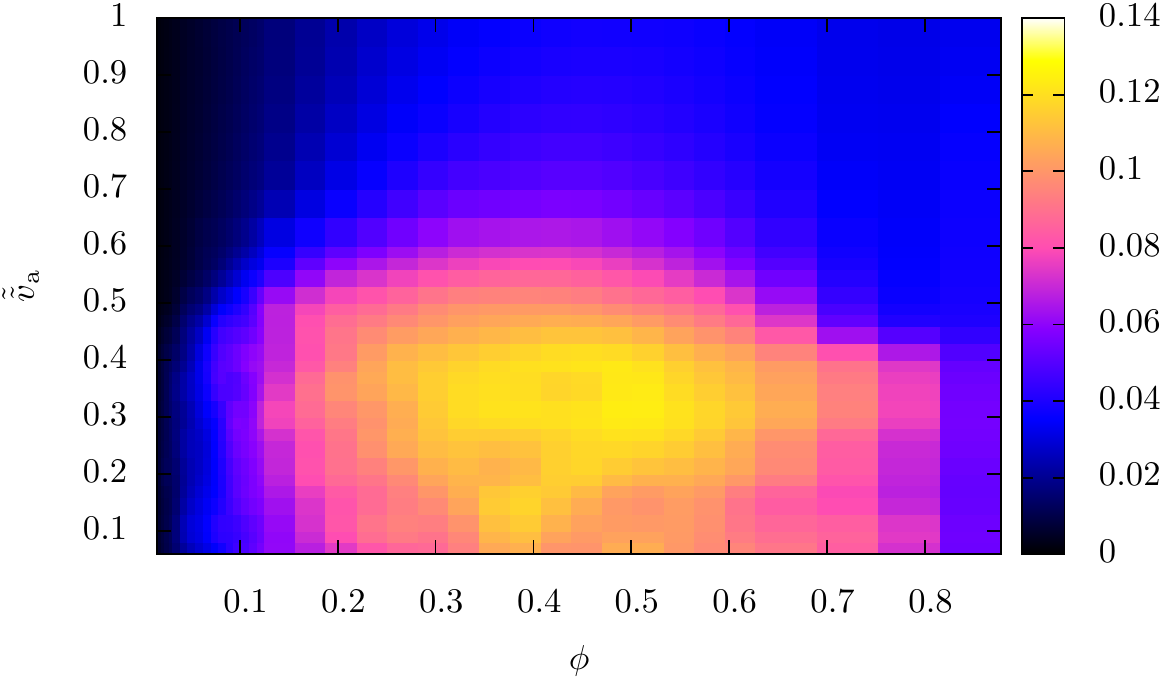}}
\subfigure[$\zeta=200$]{\includegraphics[width=8.5cm]{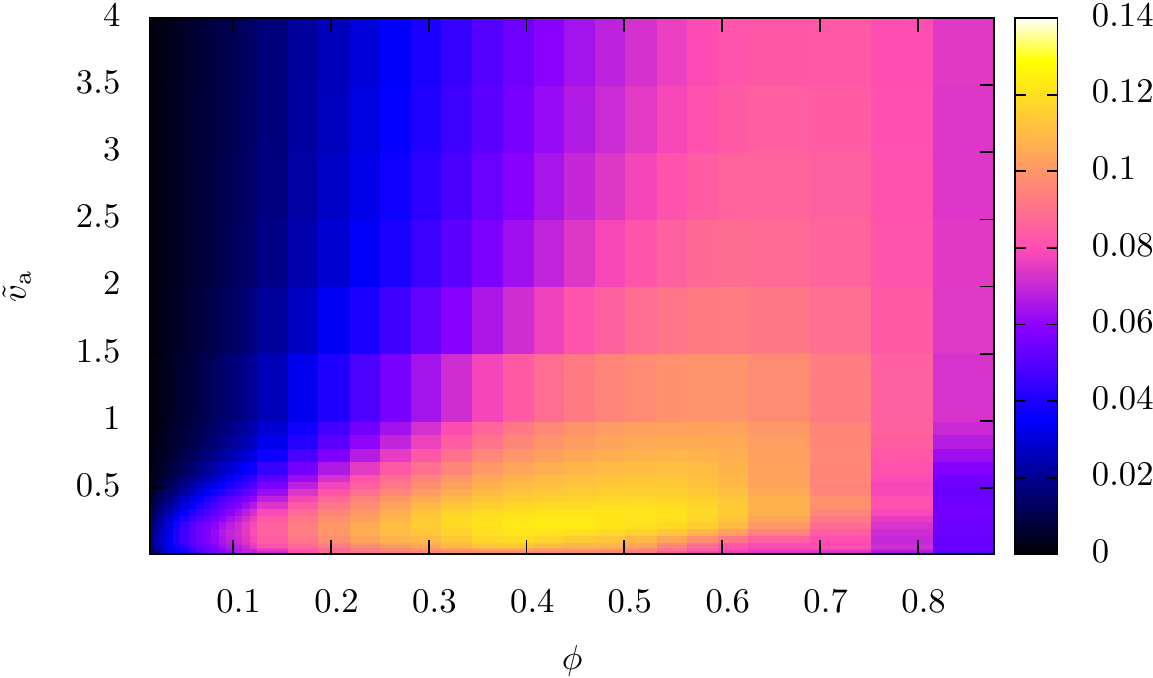}}
\caption{\label{map_dn_z_ab}Active Brownian Particles. Area fraction
  fluctuations $\delta_\phi$ for a small (upper) and large (lower)
  value of $\zeta$. Quasi-equilibrium phase separation is seen at low
  scaled active swim speed $\va$ for both values of $\zeta$. Strongly
  non-equilibrium MIPS is seen at high $\vat$ only for large $\zeta$.
  Number of boxes $G=5$.  }
\end{figure}
\begin{figure}[h] 
\centering
\subfigure[$\phi=0.125$]{\includegraphics[width=8.0cm]{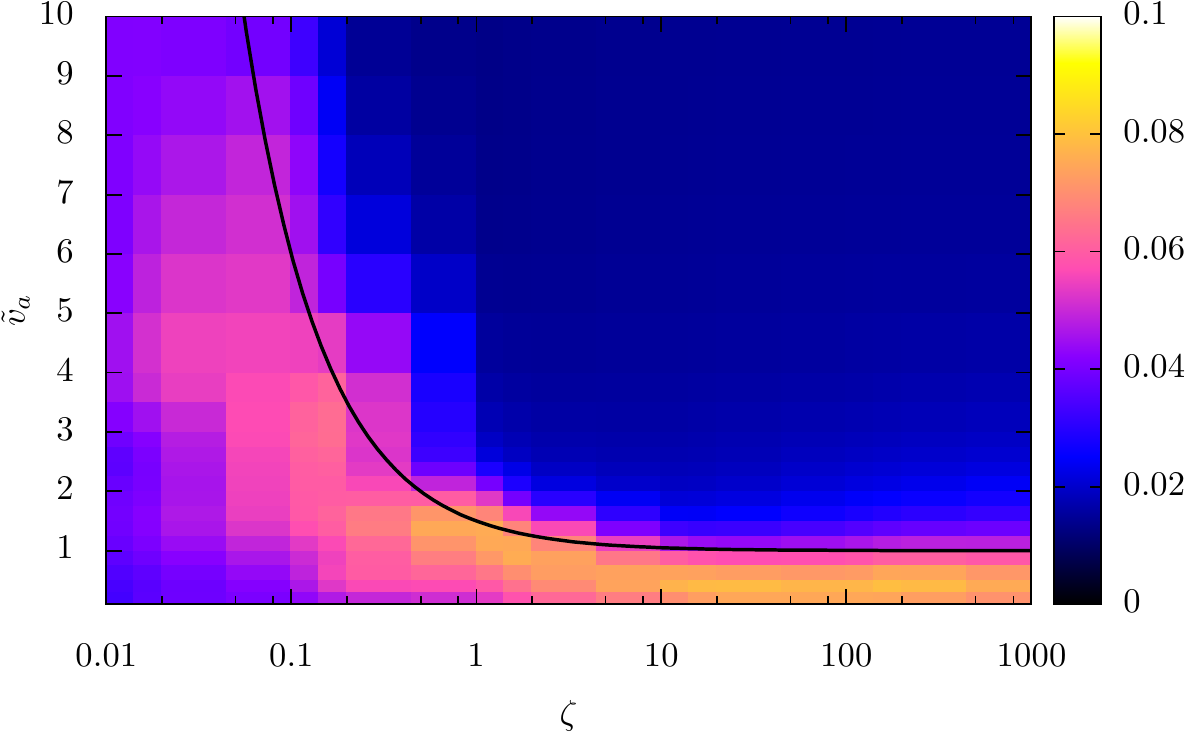}}
\subfigure[$\phi=0.75$]{\includegraphics[width=8.0cm]{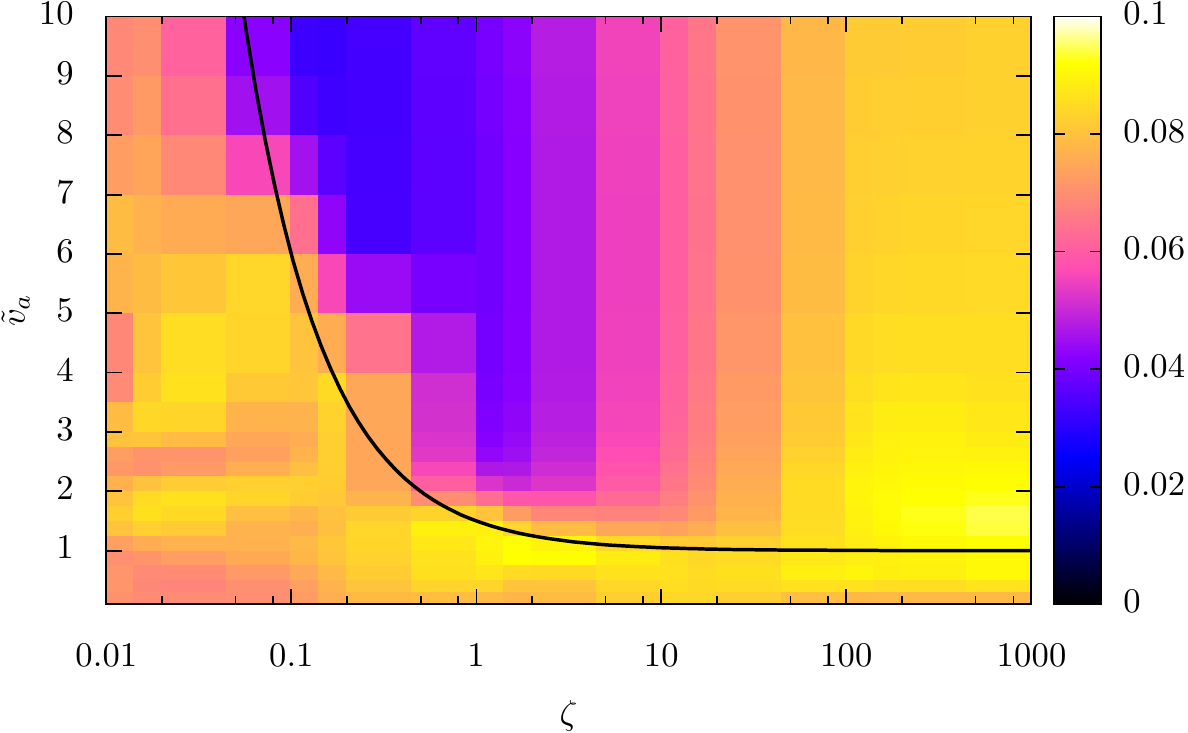}}
\caption{\label{map_dn_phi_ab}Active Brownian particles: area fraction
  fluctuations $\delta_\phi$ in the $\phi-\zeta$ plane.  The black
  line corresponds to the curve $\va\zeta/(1+\zeta)=1$. See
  Eqn.~\ref{eqn:transition} in the main text.  Number of boxes $G=5$.}
\end{figure}

Again as for the PBP, in a lower-left rectangle of the plane
$(\vatt,\phi)$ that roughly corresponds to $\phi < 0.25$ and
$\vatt<0.4$, the system doesn't reach steady state in any feasible
run-time but instead shows a characteristic power-law coarsening
process in which the typical domain size grows in time as $\bar{n}\sim
t^{\beta}$. The data are well fit by the exponent $\beta=2/3$, as in
the PBP above (data not shown).

We have established, then, that for small $\zeta$ the phase behaviour
(and coarsening kinetics) of ABP in the plane of $(\vatt,\phi)$
closely resembles that of PBP in the plane of $(\Dt,\phi)$. This
correspondence is with hindsight perhaps unsurprising, given that the
intercollisional dynamics are mainly diffusive in this regime, as in a
passive system.

We now turn to the opposite regime of large $\zeta$, where $\vat$ is
the most obvious choice for adimensionalising $\va$.
Fig.~\ref{snaps_ab_z_100} shows long-time snapshots of the system's
configuration on a grid of values of $(\vat,\phi)$ for $\zeta=200.0$.
The situation here is clearly more complicated than for the low value
of $\zeta$, with the phase diagram differing more markedly from the
equilibrium case. This is to be expected, consistent with the fact
that underlying ``microscopic'' dynamics that feeds forward to inform
the collisional dynamics is predominantly ballistic, rather than
diffusive as it was in the case either of PBPs or ABPs for small
$\zeta$.

Nonetheless, at low swim speeds the phase behaviour does appear rather
similar to that of PBPs, and so also of ABPs at small values of
$\zeta$.  Accordingly we suggest that this again corresponds to a
regime of phase separation analogous to that of the equilibrium PBPs,
arising due to the dominance of the Lennard-Jones attraction over the
active swimming for small $\va$.  However we repeat that the physics
is subtly different here: unlike the equilibrium case the
intercollisional dynamics is predominantly ballistic rather than
diffusive. In view of this, the fact that we do still see a regime of
attraction-dominated quasi-equilibrium phase separation at low $\vat$
is a non-trivial result.  On this basis, one might tentatively claim
that the ABPs' scaled swim speed has some properties resembling those
of a non-equilibrium effective temperature.

At high activity (large $\vat$), however, the phase behaviour for this
large value of $\zeta$ is markedly different from the equilibrium
case. In particular, a second regime of phase separation is evident
for values of $\phi$ exceeding about $0.4$.  Clearly, this could not
have been predicted on the grounds of any direct analogy with the
equilibrium phase diagram.  How then can it be understood?

For these large values of $\vat$ we expect the strength of swimming to
easily overcome any attraction associated with the LJ potential, such
that active LJ disks behave effectively as active hard disks in this
limit $\vat\to\infty$.  With this in mind, we recall previous work for
hard disk ABPs that demonstrated a phenomenon known as ``motility
induced phase separation'' (MIPS) ~\cite{tailleur2008,cates2010}. This
arises via a positive feedback mechanism in which particles (a)
accumulate in regions where they move more slowly, then (b) further
slow down in regions of accumulation, being impeded by overcrowding
from other particles.  Note that although part (a) of this mechanism
may seem intuitively obvious -- think of window shoppers accumulating
where they linger near a more interesting display -- it is actually a
highly non-equilibrium phenomenon: in equilibrium colloids the local
density depends only on the static potential and not on any dynamical
rules.  Informed by this, we interpret this second region of phase
separation, seen at high $\vat$ for large $\zeta$, as one of MIPS.

The absence of MIPS at high $\vat$ for small values of $\zeta$, as in
Fig.~\ref{snaps_ab_z_.2}, is then further
understood~\cite{PhysRevLett.112.118101,fielding14} by realising that
part (b) of the feedback mechanism just described contains a mean
field assumption that fails for small $\zeta$: in order to experience
a slowing down of its run speed, a particle must collide many times
during any such run before changing direction.  This assumption breaks
down for small values $\zeta$, where particles reorient very quickly
on the timescales of interparticle collisions.

This difference between the phase behaviour of ABPs for small and
large values of $\zeta$ is characterised at a glance by colour maps of
the particle number fluctuations in Fig.~\ref{map_dn_z_ab}: MIPS is
apparent for large $\zeta$ (bottom panel) for $\phi\gae 0.5$ and
$\vat\gae 1.0$; and absent for small $\zeta$ (top panel).

The dependence on $\zeta$ is shown more fully in
Fig.~\ref{map_dn_phi_ab}.  For the small area fraction $\phi=0.125$,
no MIPS is expected for any value of $\zeta$ (recall
Fig.~\ref{map_dn_z_ab}) and we see only quasi-equilibrium phase
separation driven by the Lennard-Jones attractions. As can be seen,
this arises for $\vat <O(1)$ at large $\zeta$, consistent with the
expectation that $\vat=\va R\gamma/\epsilon$ is the relevant scaling
variable in this regime.  In contrast, for low $\zeta$ it arises
instead for $\vatt<O(1)$, consistent with the expectation that
$\vatt=\va^2\gamma/\Dr\epsilon$ is the relevant scaling variable in
this regime.  
Noting that $\vatt=\vat\zeta$, we find finally
that
\be
\label{eqn:transition}
\frac{\vat\zeta}{1+\zeta}<1
\ee
gives a good fit to the regime of attraction-dominated phase
coexistence for all values of $\zeta$, recovering $\vatt<1$ for
$\zeta\ll 1$ and $\vat<1$ for $\zeta\gg 1$. 

For the larger area fraction $\phi=0.75$ in Fig.~\ref{map_dn_phi_ab}
we find the same picture with regards attraction-dominated phase
separation, with the transition again given by
Eqn.~\ref{eqn:transition}. In addition to this we find a regime of
MIPS for large values of $\zeta\gae 10.0$ and $\vat \gae 1.0$.
Interestingly, these regimes of equilibrium-like attraction-dominated
phase separation and MIPS appear to merge with each other smoothly,
with a gradual cross over regime in which both mechanisms
contribute. We note that re-entrance between two regimes of phase
separation was also observed in simulations of attractive ABPs with
the conventional LJ potential in Ref.~\cite{redner13b}. However that
work considered only the regime of small rotational diffusion
coefficient (of thermal origin, and so related to the Brownian
diffusion coefficient via a factor $3/R^2$) and therefore necessarily
saw a regime with MIPS. The present study goes beyond the findings
of~\cite{redner13b} in showing MIPS to be suppressed with increasing
rotational diffusion. (RB11/15)In Ref.~\cite{prymidis15}, a crossover from
equilibrium-like phase behaviour to a non-equilibrium percolating
network phase was seen in a system of active Brownian Lennard Jones
particles as function of decreasing rotational diffusion
coefficient. The percolating network phase in that study might well be
the counterpart of the MIPS seen here, though further work would be
needed to elucidate further any connection.

\subsection{Hydrodynamic squirmers}
\label{sec:HSresults}

In the previous section, we explored the phase behaviour of active
Brownian particles (ABPs), which combine ballistic swimming with
stochastic angular reorientation that is intended to mimic, in
continuous-time counterpart, discrete tumbling events in some species
of bacteria, or scattering off other particles due to hydrodynamic
interactions.  Hydrodynamic interactions are not, however, explicitly
accounted for in ABPs. In this section we turn to address that
shortcoming by studying a suspension of squirmers that do interact
hydrodynamically.

We start with a recap of the parameters that characterise squirmers
and the appropriate adimensional form of these, as first introduced in
Sec.~\ref{sec:parameters} above. Indeed to set these in a proper
context, we recall them alongside the (by now) familiar parameters of
the PBP and ABPs.

As always we take as fixed the number of particles $N=242$ and the
lengthscales $\sigma_{\rm h}$ and $\sigma_{\rm LJ}$ that set the size
of the potential shell relative to the hydrodynamic radius $R$. For
the PBPs the remaining five parameters are then the particle radius
$R$, the potential energy $\epsilon$, the drag coefficient $\gamma$,
the area fraction $\phi$ and the thermal translational diffusion
coefficient $D$.  Choosing units of mass, length and time left two
remaining dimensionless parameters: $\Dt=D\gamma/\epsilon=\kB
T/\epsilon$ and $\phi$. In Sec.~\ref{sec:PBPresults} we explored the
phase behaviour of our modified Lennard-Jones particles in this plane
of $(\Dt,\phi)$.
\begin{figure*} 
\centering
\includegraphics[width=16.0cm]{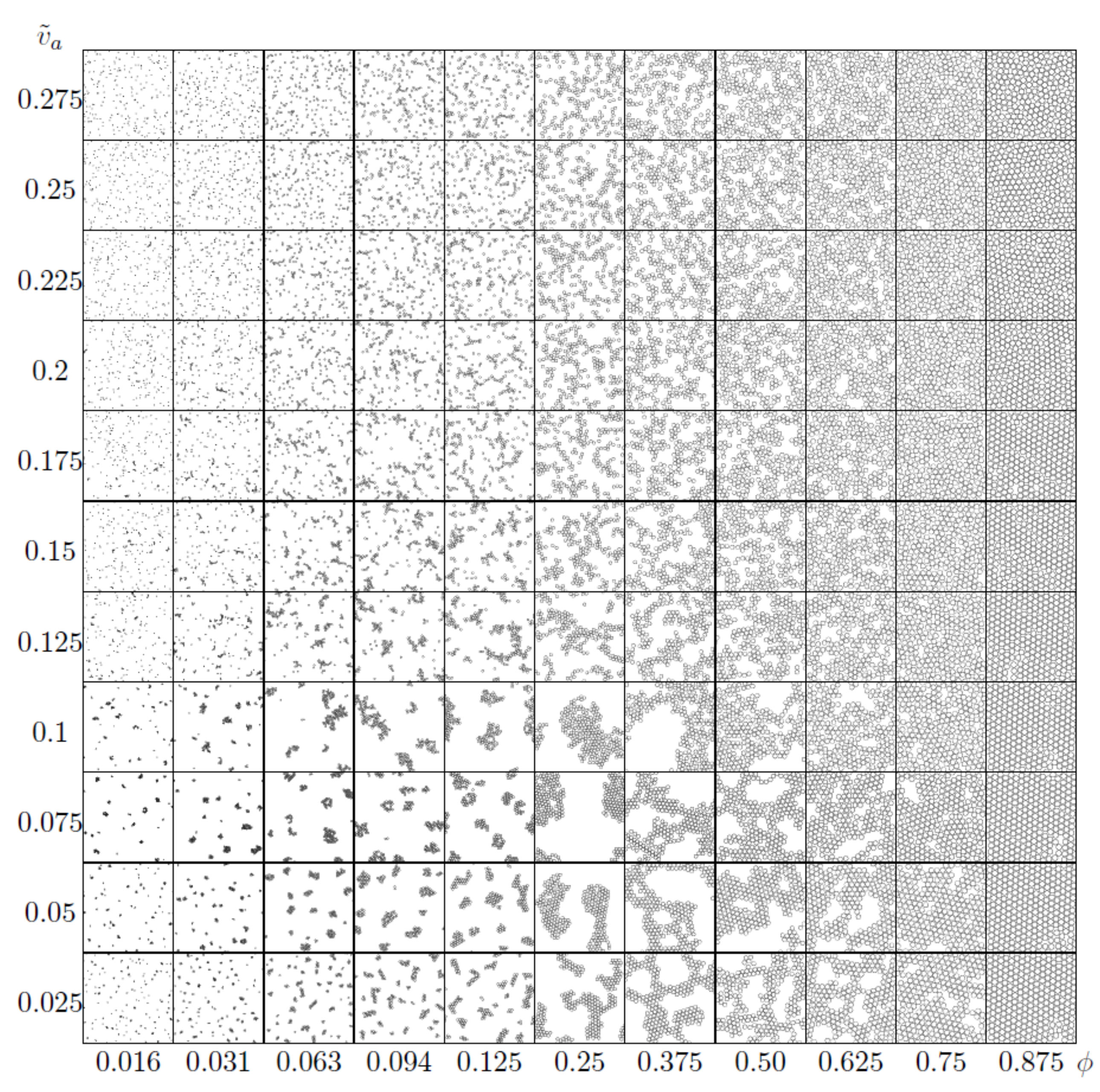}
\caption{\label{squirmconfigmap} Hydrodynamic squirmers, $\beta=0.0$.
  Snapshots of the system's configuration on a grid of values of the
  scaled active swim speed $\vat$ and effective area fraction $\phi$.
  Each snapshot is taken at a long time after the system was
  initialised at $t=0$ in a random state as described in
  Sec.~\ref{sec:initial}.  In the rectangle $\vat < 0.15$ and
  $\phi<0.1$ the system fails to reach steady state in any feasible
  runtime and instead shows slow domain coarsening. All other
  snapshots shown are from a system in statistically steady state.
  Note that the scale is nonlinear at the largest $\vat$ and at low
  $\phi$.}
\end{figure*}
For the ABPs the remaining parameters were the particle radius $R$, the
potential energy $\epsilon$, the drag coefficient $\gamma$, the area
fraction $\phi$, the swim speed $\va$ and the rotational diffusion
coefficient $\Dr$. In dimensionless form these gave three parameters,
which we explored numerically in the previous section. Two of these
were the area fraction $\phi$ and the scaled inverse rotational
diffusion coefficient $\zeta=\va/\Dr R$, which characterises the time
taken for the angular decorrelation of swim direction relative to the
typical time interval between particle collisions.  The third was the
swim speed $\va$, for which we argued $\vatt=\va^2\gamma/\Dr\epsilon$
to be an appropriate adimensional form in the regime of small $\zeta$
where the inter-collisional dynamics are diffusion dominated, and
$\vat=\va R\gamma/\epsilon$ at high $\zeta$, where the
inter-collisional dynamics are predominantly ballistic.  Given that
$\vatt=\vat\zeta$ these two coincide for values of $\zeta=O(1)$, as
desired.

For the hydrodynamic squirmers to which we now turn the relevant
parameters are the particle radius $R$, the potential depth
$\epsilon$, the solvent viscosity $\eta$, the area fraction $\phi$,
the swim speed $\va=B_1/2$ and the relative stresslet strength
$\beta=B_2/B_1$.  In dimensionless form we then have the scaled swim
speed $\vat=\va\eta R^2/\epsilon$, the area fraction $\phi$, and the
stresslet strength $\beta$. We shall show in what follows that the
effects of $\beta$ on phase behaviour are in fact relatively mild.
Accordingly we shall focus first, and mainly, on phase behaviour as a
function of $\vat$ and $\phi$, before returning at the end of the
section to comment on the effects of varying $\beta$.

Excluding $\beta$ for the moment, then, we note the number of
important parameters ($\vat,\phi$) for the squirmers to be one fewer
than those for the ABPs, ($\vat$, $\phi$, $\zeta$) (or
($\vatt,\phi,\zeta$)). The reason for this is as follows.  Whereas the
ABPs have externally imposed reorientation dynamics with a rotational
diffusion coefficient $\Dr$, adimensionalised as $\zeta$, the
squirmers have no reorientation dynamics imposed upfront.  Instead,
reorientation events emerge naturally as a result of the hydrodynamic
interactions that determine the way in which the squirmers scatter off
each other (at low area fraction) or slither round each other (at high
area fraction).  In consequence we expect an effective ratio
$\zeta_{\rm eff}\equiv \tauo/\tauc$ of reorientation time to
scattering time to emerge naturally from the squirmers' many body
hydrodynamics, instead of existing as a free parameter to be imposed
at the outset in the simulations.

Having identified the relevant parameters, a natural question is then
whether the phase behaviour of the squirmers in the plane of
$(\vat=\va\eta R^2/\epsilon, \phi)$, with whatever value of
$\zeta_{\rm eff}$ emerges naturally from the hydrodynamic simulations,
mimics that of the ABPs in the plane $(\vat=\va \gamma
R/\epsilon,\phi)$ for a corresponding value of $\zeta$, as imposed for
the ABPs. One can of course also further ask whether it further mimics
that of PBPs in the plane of $(\Dt,\phi)$.

Clearly, the dimensionless forms for $\va$ identified above suggest
that we might look for a possible correspondence between ABPs and
squirmers (after matching  the imposed $\zeta$ for the ABPs with the emergent
$\zeta_{\rm eff}$ for the squirmers) upon setting the ABPs' drag
coefficient $\gamma=\eta R$, where $\eta$ is the solvent viscosity for
the squirmer simulations.  However it should be noted that this
expression should also contain a prefactor, which is unknown. The
specific case of a sphere dragged by a force monopole through an
infinite domain of fluid in 3D gives $\gamma=6\pi\eta R$.  However we
are concerned here neither with force monopoles nor with 3D and the
prefactor remains unknown, even if we might reasonably expect it to be
$O(1)$. With this preamble in mind we now present our results. We start by
taking a fixed $\beta=0.0$ and shall return at the end of the section
to comment on the effect of varying $\beta$.
Fig.~\ref{squirmconfigmap} shows snapshots of the system's
configuration on a grid of values of $(\vat,\phi)$. Each snapshot
corresponds to the longest run time that is feasibly accessible
computationally.  Most represent a statistically steady state, except
in a region towards the lower left of the $(\vat,\phi)$ plane,
typically for values of $\vat<0.15$ and $\phi<0.1$.  Here the
coarsening dynamics of the aggregated domains, which we shall discuss
in more detail below, is sufficiently slow that the system never
attains steady state in this regime in any reasonable run
time. Despite this we are confident that running for even longer times
would not affect the qualitative features of the system's state, apart
from giving larger domains.

\begin{figure}[h!] 
\centering
\subfigure[(a)]{\includegraphics[width=8.cm]{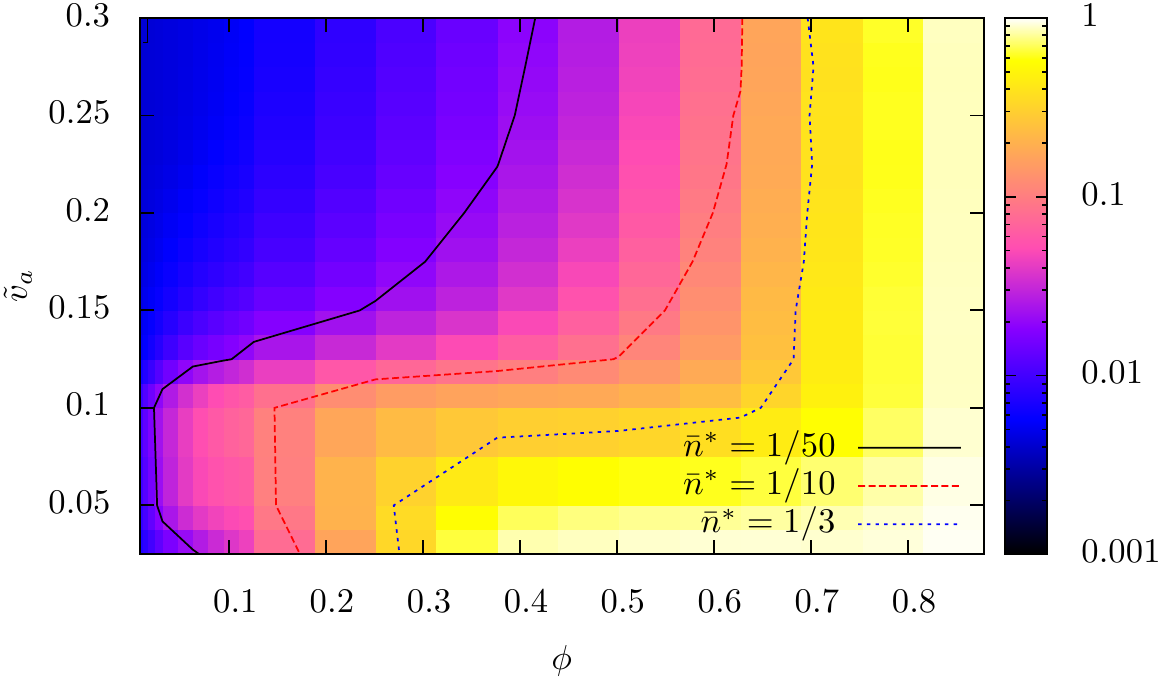}}
\subfigure[(b)]{\includegraphics[width=8.cm]{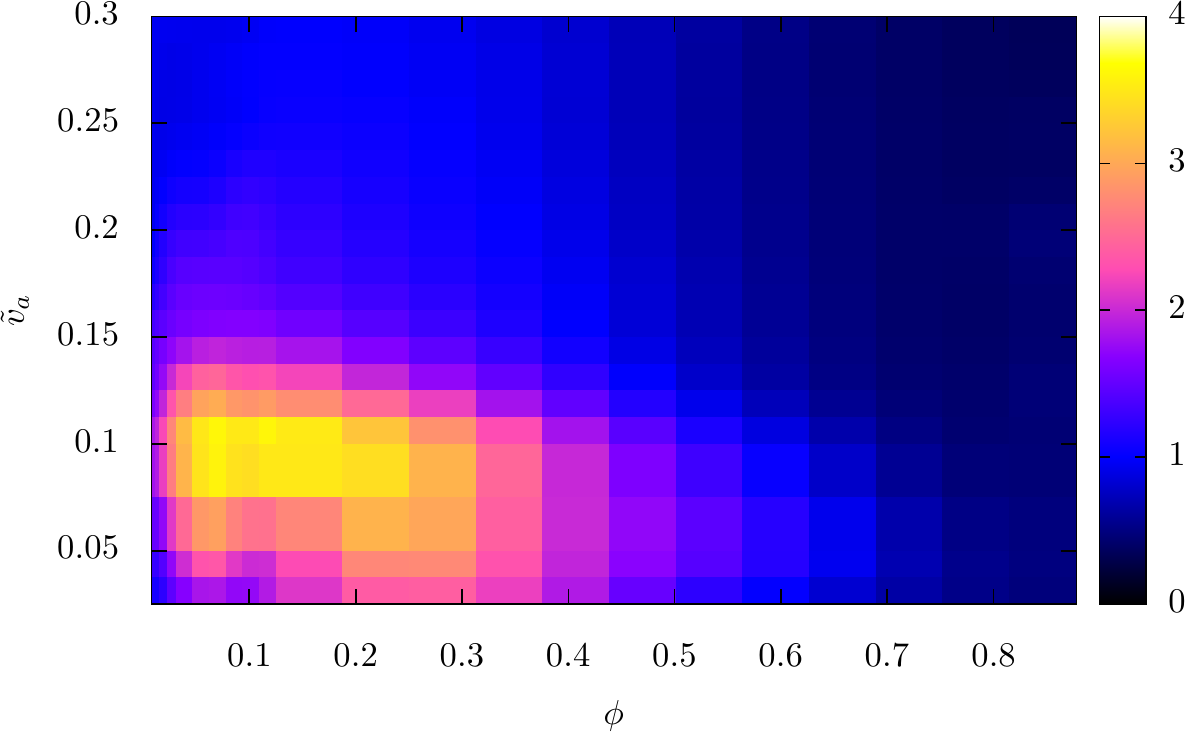}}
\subfigure[(c)]{\includegraphics[width=8.cm]{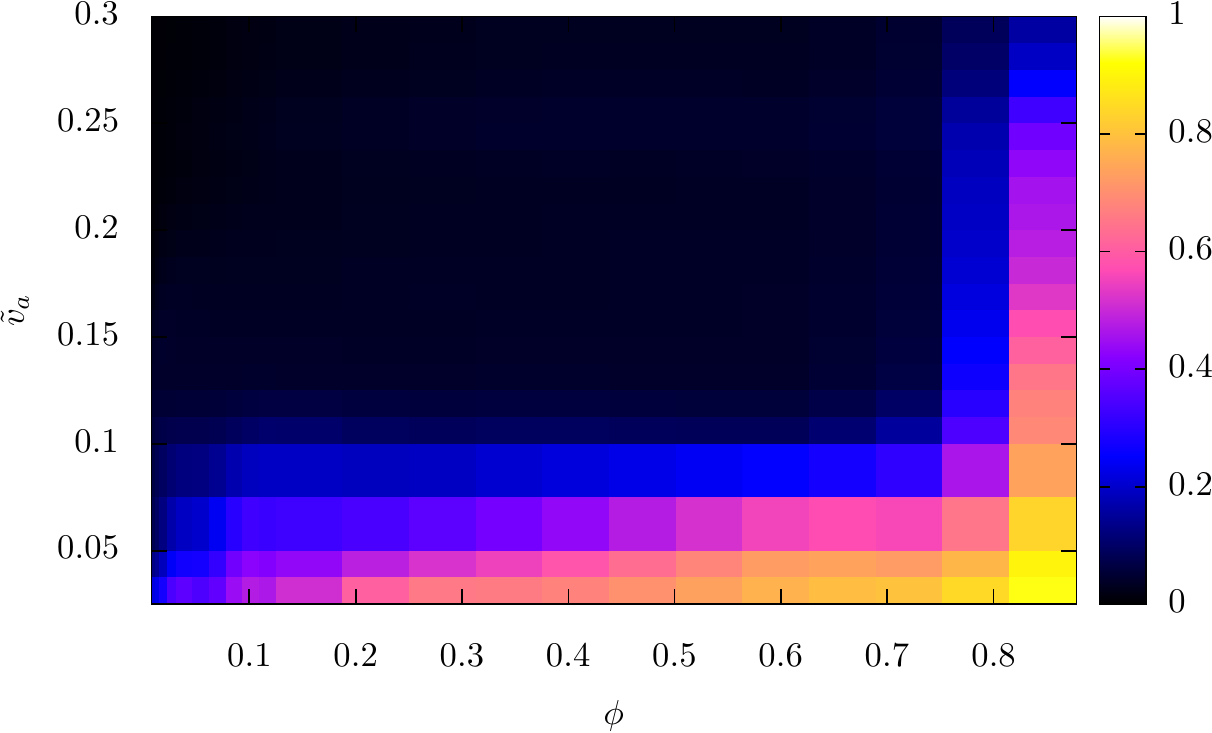}} 
\caption{\label{map_n0npc_s} Phase diagram of hydrodynamic squirmers
  mapped using (a) normalised mean number of particles per cluster
  $\bar{n^*}=\bar{n}/N$, (b) particle number fluctuations
  $\delta_{N}/\sqrt{\bar{N}}$ for $G=5$ and c)
  hexatic order parameter $\Psi_6$.}
\end{figure}

Some quantitative measures of this phase diagram are shown in
Fig.~\ref{map_n0npc_s}. Panel a) gives the mean number of particles
per cluster $\bar{n}$, normalised by the total number of particles $N$
to give $\bar{n}^*=\bar{n}/N$. Recall that a single system-spanning
condensate would have $\bar{n}^*=1$, while a gas phase with no
clusters would have $\bar{n}^*=1/N$, which is small for large $N$.
Panel b) shows the particle number fluctuations
$\delta_N/\sqrt{\bar{N}}$ measured by dividing the simulation box into
$G\times G$ subboxes where $G=5$. Subpanel c) shows the hexatic order
parameter $\Psi_6$.  Recall that regions of high $\bar{n}^*$ and
$\delta_N/\sqrt{\bar{N}}$ indicate clustering or phase separation, while
regions of high $\Psi_6$ indicate a high degree of solid-like
crystalline ordering.

Comparing these snapshots and colourmaps for the squirmers
(Figs.~\ref{squirmconfigmap} and~\ref{map_n0npc_s}) with their
counterparts for the PBPs (Figs.~\ref{passiveconfigmap}
and~\ref{fig:PBPphasediag}) reveals that the overall phase behaviour
of the squirmers loosely mirrors that of the PBPs, with the scaled
swim speed $\vat$ playing a role somewhat analogous to that of the
scaled diffusion coefficient $\Dt$ of the PBPs.  In particular,
attraction-dominated phase separation and hexatic ordering appears
evident at low $\vat$ (corresponding to low $\Dt$ for the PBPs),
giving way to disorder and homogeneity at high $\vat$ (high $\Dt$ for
the PBPs) where the particles' activity (thermal jostling for the
PBPs) is sufficient to overcome the attractive effects of the
potential. Solid-like ordering is also evident in a column of values
of $\vat$ for densely crowded systems with $\phi>0.75$.

However there are also important differences. For example, in the
squirmers hexatic ordering (Fig.~\ref{map_n0npc_s}c) is only strongly
apparent at the lowest values of $\vat$ for which appreciable area
fraction fluctuations are present (Fig.~\ref{map_n0npc_s}b),
reflecting the fact that the evidently phase separated snapshots in
Fig.~\ref{squirmconfigmap} only show solid-like ordering at the lowest
$\va$. In contrast, for the PBPs hexatic order
(Fig.~\ref{fig:PBPphasediag}c) persists to higher values of $\Dt$
relative to those for which appreciable phase separation is evident
(Fig.~\ref{fig:PBPphasediag}b).  The same is true of the column of
hexatic ordering seen at high $\phi$: this gradually fades with
increasing $\vat$ in the squirmers, whereas it apparently persists to
indefinitely high $\Dt$ for the PBPs in Fig.~\ref{fig:PBPphasediag}c.

This gradual decay of hexatic order with increasing $\vat$ suggests
that the attraction-dominated phase ordering is disrupted more readily
by the ballistic intercollisional swimming of the squirmers with
increasing $\vat$ than it is by the diffusive intercollisional
dynamics of the PBP with increasing $\Dt$. This is consistent with
the fact that these coherently moving ``living crystals'' seen in the
regime of clustering at low $\vat$ in Fig.~\ref{squirmconfigmap}, are
much more dynamic and mobile entities than their equilibrium
counterparts in Fig.~\ref{passiveconfigmap}. The mean squared
displacement (MSD) of particles in these clusters (data not shown)
have an early-time ballistic regime crossing over into a later-time
diffusive regime, even in crowded systems. In contrast, the
counterpart MSDs for crowded PBPs at low $\Dt$ have a
slower-than-diffusive (``caged'') regime at early times crossing over
into diffusive dynamics later on (data not shown).  Movies of these
clusters can be found online$^\dag$ and are seen to resemble those
observed experimentally in active colloids in
Refs.~\cite{buttinoni13,palacci13,PhysRevX.5.011004}, named as
``living crystals''.

Another notable difference is that stringlike clusters remain clearly
evident in the squirmer snapshots even at high $\vat$
(Fig.~\ref{squirmconfigmap}), compared with the more complete return
to homogeneity at high $\Dt$ in the PBPs
(Fig.~\ref{passiveconfigmap}).  This is consistent with the existence
of a larger window between the lines of $\bar{n}^*=1/50$ and $1/10$ in
Fig.~\ref{map_n0npc_s}a compared to its equilibrium counterpart in
Fig.~\ref{fig:PBPphasediag}a, indicating a larger regime of clustering
in the squirmers than in the PBPs.  Likewise the area fraction
fluctuations decay more slowly as a function of $\vat$ in
Fig.~\ref{map_n0npc_s} than they do as a function of $\Dt$ in
Fig.~\ref{fig:PBPphasediag}.

Clearly, this slight clustering of squirmers in the regime of high
$\vat$ has no counterpart in the equilibrium phase diagram of the PBPs
at high $\Dt$. It is, however, reminiscent of the phase behaviour of
ABPs for low to moderate values of $\zeta$, where a slight clustering
arises at high $\vat$, which we interpreted as a precursor to the
onset of true MIPS at high $\zeta$.

\begin{figure}[h]
\centering
\includegraphics[width=8.0cm]{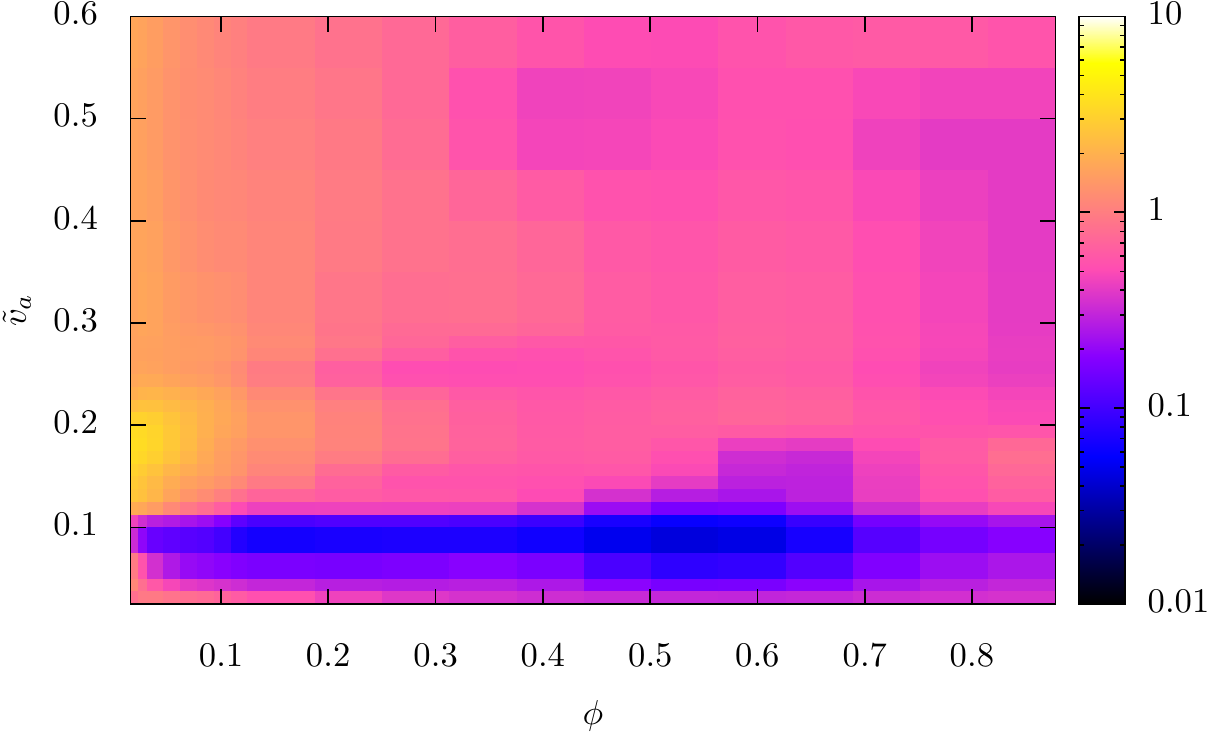} 
\caption{\label{emergentzeta}Ratio between the characteristic
  reorientation time $\tauo$ and collision time $\tauc$ for polar
  squirmers, giving the emergent effective ratio $\zeta_{\rm eff}$. }
\end{figure}

Tentatively, then, we interpret this tendency of the squirmers to form
stringlike clusters at high $\vat$ as a counterpart of the slight
clustering seen in the ABPs for modest imposed values of $\zeta=O(1)$.
Recalling that $\zeta$ corresponds to the ratio of reorientation time
and (estimated) collision time $\tauo/\tauc$ that is imposed upfront
in the ABP simulations, if this interpretation is correct we might
then reasonably anticipate that the corresponding effective ratio
$\zeta_{\rm eff}\equiv\tauo/\tauc$ that {\em emerges} from the
squirmer simulations should likewise be $O(1)$.  This prediction of an
emergent $\zeta=O(1)$ was indeed confirmed (for the case of hard disk
squirmers) in Ref.~\cite{fielding14}.  It is furthermore consistent
with earlier observations that whenever two squirmers scatter off each
other hydrodynamically they also suffer an $O(1)$ change in their swim
directions~\cite{scatter}. To reinforce this argument we extracted
from our squirmer simulations this emergent effective $\zeta_{\rm
  eff}$, measuring the characteristic times $\tauo$ and $\tauc$ in
the way defined in Ref.~\cite{fielding14}. As can be seen in
Fig.~\ref{emergentzeta}, we indeed recover $O(1)$ (or smaller) values
for the effective emergent $\zeta_{\rm eff}$ for all the mid-high area
fractions, which would be the ones potentially inside the MIPS regime.

We therefore suggest that a slight tendency to clustering seen at high
$\vat$ in the squirmers is the signature of a nearby MIPS, the full
effects of which have been mitigated by the fact that the effective
emergent value of $\zeta_{\rm eff}$ for the squirmers is only $O(1)$,
rather than the larger value of $\zeta$ needed to see full MIPS in the
ABPs.  The underlying physics here is that, as just noted, squirmers
change their orientations each time they scatter off each other
hydrodynamically, thereby giving an effective $\zeta_{\rm eff}=O(1)$.
In contrast for the ABPs we can artificially tune $\zeta$ to be
arbitrarily large simply by setting a smaller $\Dr$ in the
simulations, thereby inducing MIPS.

As noted above, this suppression of MIPS by hydrodynamics was first
discussed in the context of purely repulsive particles in
Ref.~\cite{fielding14}.  A contribution of the present work
has been to establish that MIPS is also suppressed by hydrodynamics in
systems of attractive particles, despite any intuition one might have
held upfront that attractive interactions might enhance the crowding
mechanism of MIPS.  The situation could however be different in
systems such as in Janus particles~\cite{PhysRevE.89.050303} where the
interaction potential is orientation-dependent and might be expected
to increase the angular decorrelation time $\tauo$, thereby increasing
the tendency towards MIPS.

Although we have demonstrated the suppression of MIPS by hydrodynamics
only in this athermal limit of $\Dr=0$ and $D=0$ for which we simulate
the squirmers, it was in this limit that MIPS was found to be most
pronounced in systems of ABPs without hydrodynamics in
Ref.~\cite{redner13a}.  This gives a strong indication that MIPS
should remain suppressed even at non-zero temperature, $\Dr\neq 0$ and
$D\neq 0$. Indeed, any extra thermal contribution to $\Dr$ would be
expected only further to decrease the ratio of $\zeta_{\rm
  eff}=\tauo/\tauc$, thereby further suppressing MIPS. The same
argument would apply to the explicit inclusion of (athermal) tumbling
events in the simulations.

To summarise, we suggest that the phase behaviour of attractive
squirmers in the plane of $(\vat,\phi)$ can be understood in terms of
two different mechanisms that smoothly blend into each other. The
first is a regime of attraction-dominated phase separation at small
$\vat$, which we suggest is analogous to the equilibrium phase
separation seen in PBPs at small $\Dt$, although more easily disturbed
by the ballistic intercollisional dynamics of the squirmers as $\vat$
increases.  The second is a regime of string-like particle clustering
at high $\vat$, which we have interpreted in terms of a nearby MIPS
that has been mitigated by the fact that $\zeta_{\rm eff}=\tauo/\tauc$
is automatically $O(1)$ for squirmers, which reorient each time they
scatter off each other.

\begin{figure}[h!]
\centering
\subfigure[$\beta=0$]{\includegraphics[width=8.cm]{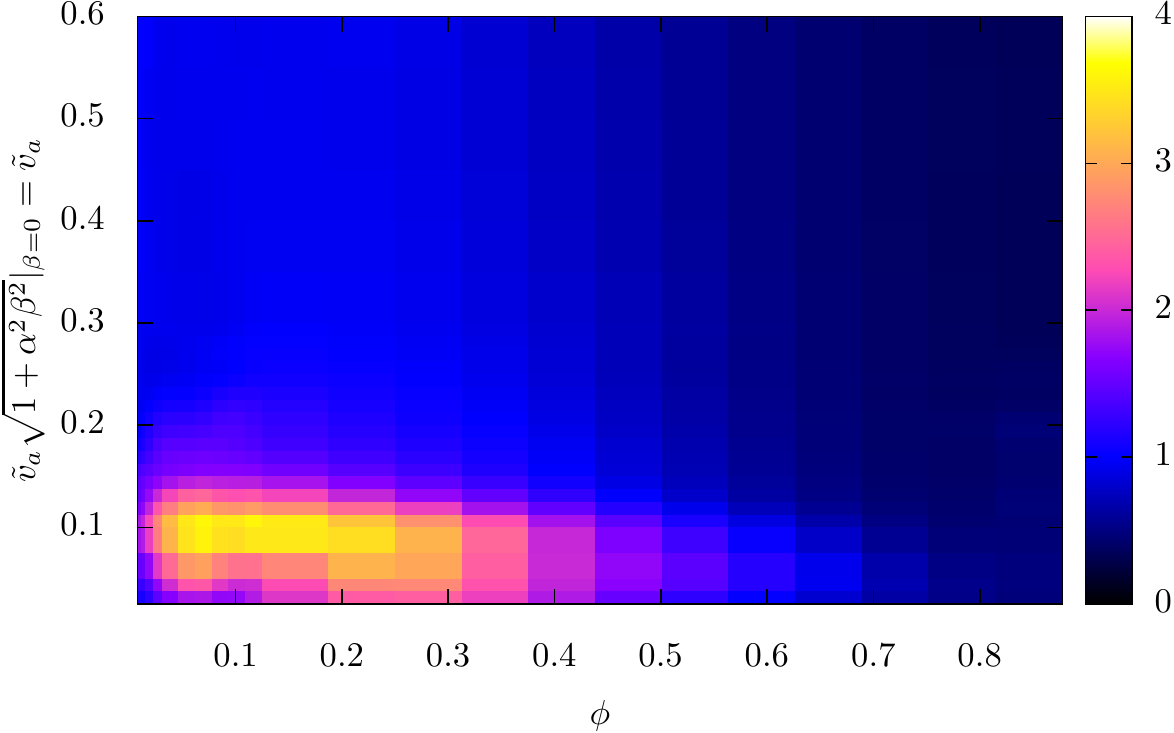}}
\subfigure[$\beta=5$]{\includegraphics[width=8.cm]{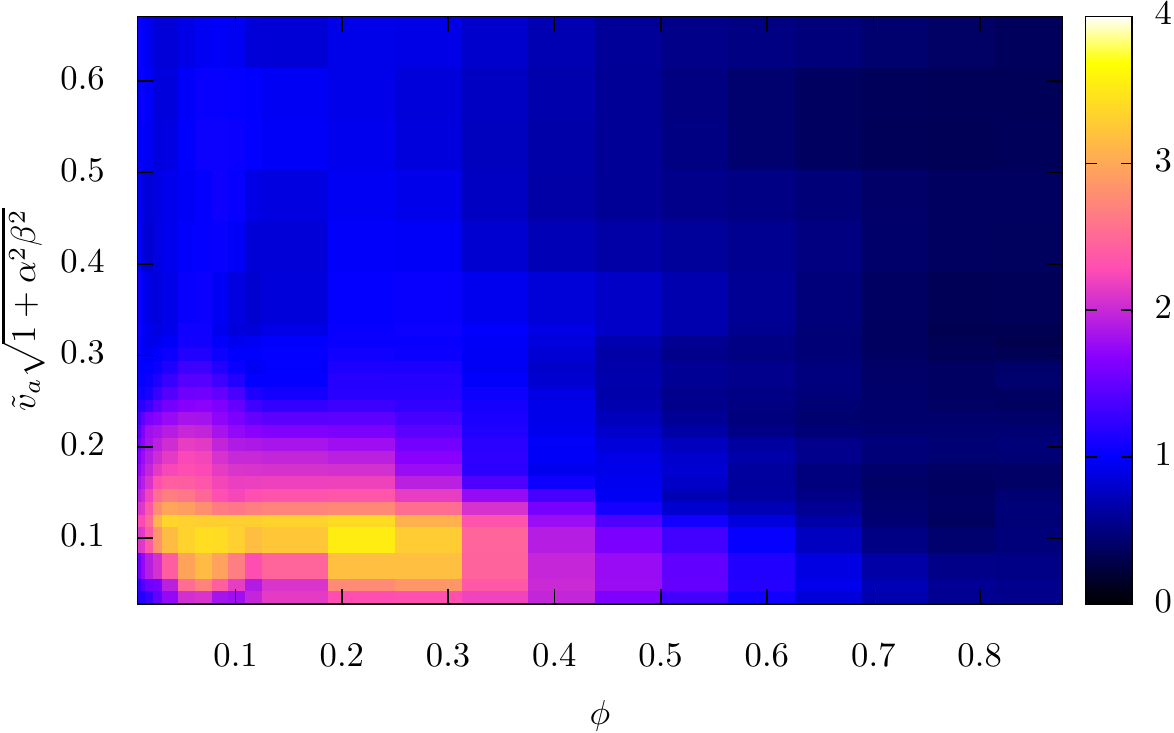}}
\subfigure[$\beta=\infty$]{\includegraphics[width=8.cm]{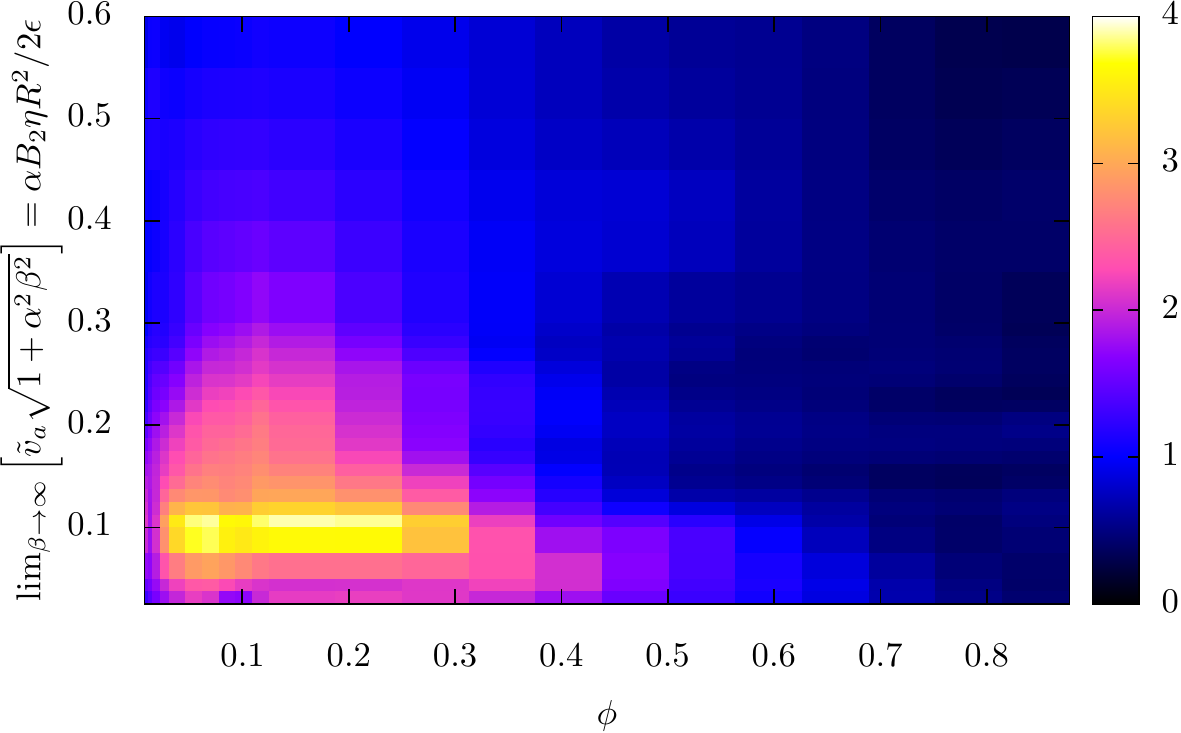}}
\caption{\label{sphisquirmersrescaled}Hydrodynamic squirmers. Area
  fraction fluctuations $\delta \phi$ obtained on dividing the
  simulation box into $G\times G$ subboxes with $G=5$.}
\end{figure}

So far, we have presented results for the hydrodynamic squirmers all
for a single value of the stresslet parameter $\beta \equiv B_2/B_1 =
0.0$. We shall now briefly consider the effects on phase behaviour of
varying $\beta$.  Recall that a single particle in an infinite medium
swims at a speed $\va=B_1/2$, which is set only by $B_1$.  The
parameter $B_2$ instead measures an additional apolar contribution to
the flow field that establishes round a (single, isolated) swimmer,
and gives no contribution to the single-particle swim speed.  In terms
of the ratio $\beta=B_2/B_1$, therefore, the regime $|\beta|\ll 1$
corresponds to strongly polar swimmers while $|\beta| \gg 1$
corresponds to apolar ``shakers''.

\begin{figure}[h] 
\centering
\includegraphics[width=8.0cm]{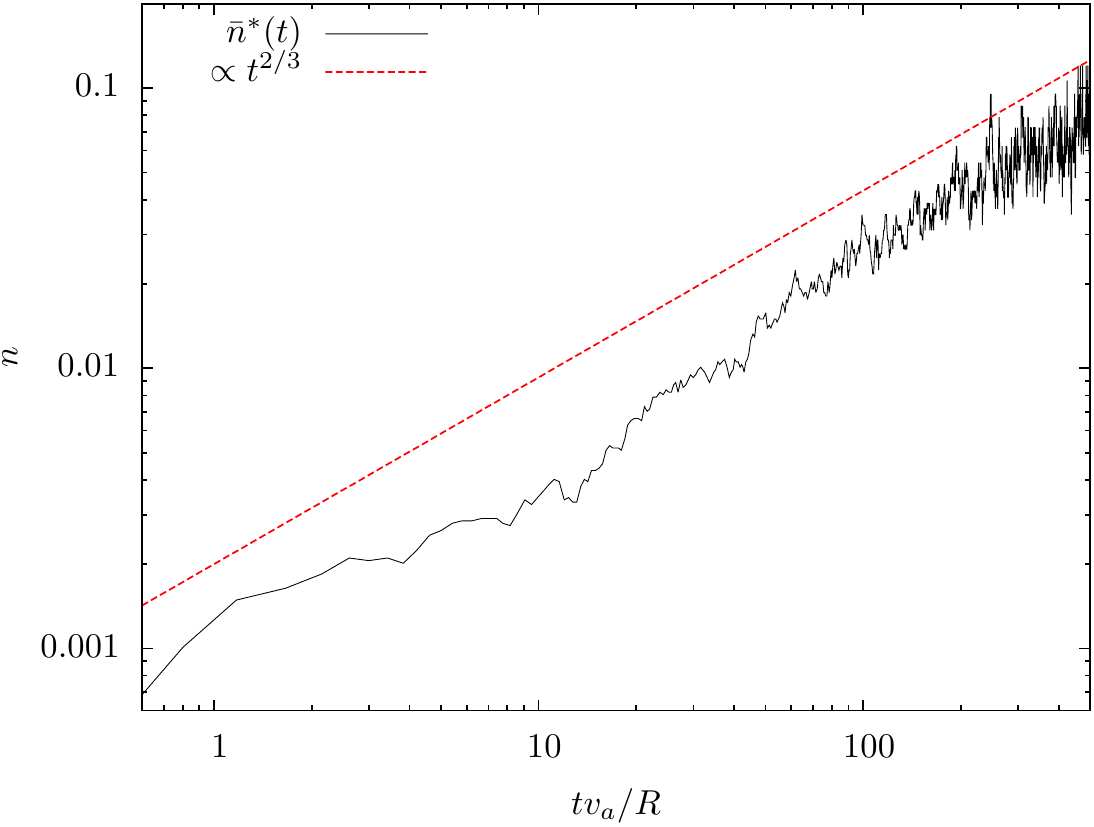}
\caption{\label{npctsquirm} Hydrodynamic squirmers, $\beta=0.0$.
  Temporal evolution of the number of particles per cluster,
  $\bar{n}^*(t)$ (minus the initial contribution $\bar{n}^*(t=0)$).
  Time is shown in units of the characteristic time $v_a/R$ that an
  isolated particle would take to displace one particle radius. The
  dotted line shows the power law $t^{2/3}$.}
\end{figure}

Although $B_2$ has no effect on the swim speed of a single particle in
an infinite medium, in the case of swimmers crowded together we might
anticipate that the additional apolar flows governed by $B_2$ do lead
to higher swim speeds for all the squirmers collectively, as the
result of hydrodynamic interactions between them: the apolar
contribution to the flow field round one squirmer disturbs
neighbouring squirmers, causing them to move slightly faster.

In view of this, we suggest that the adimensionalised measure of
activity $\vat$ on the vertical axis of the phase diagram should now
be generalised from the bare quantity $\vat=B_1\eta
R^2/2\epsilon=\va\eta R^2/\epsilon$ used so far to a quantity that
more fairly reflects the additional contribution of $B_2$ to the
overall collective activity. Accordingly we combine $B_1$ and $B_2$ as
$\sqrt{B_1^2+ \alpha^2 B_2^2}$, and define an updated adimensional
$\vat=\va\sqrt{1+\alpha^2 \beta^2}\eta R^2/\epsilon$. (This reduces as
required to the quantity $\vat=\va\eta R^2/\epsilon$ used to represent
the $\beta=0$ results above.) A key question then is whether the
squirmers' phase behaviour in the plane of (this newly defined) $\vat$
and $\phi$ depends in any significant way on $\beta$.

The parameter $\alpha$, which is defined by the formulae just given,
deserves some discussion. In general it should depend on the area
fraction $\phi$, because the degree to which the apolar components of
the flow field round each particle will cause neighbouring particles
to move via hydrodynamic interactions will depend also on the particle
separation and so on $\phi$, consistent with the fact that $B_2$ has
no effect on swimming dynamics in the single particle limit $\phi\to
0$, where only $B_1$ matters. However we shall ignore this
complication in what follows and simply set $\alpha$ constant
throughout.

Fig.~\ref{sphisquirmersrescaled} shows the phase behaviour of the
squirmers, as measured by the particle number fluctuations, in this
plane of $\vat=\va\sqrt{1+\alpha^2 \beta^2}\eta R^2/\epsilon$ and
$\phi$ for a value $\alpha=0.1$. For this particular chosen value of
$\alpha$, the main features of the phase diagram are preserved in both
magnitude and location even as $\beta$ is varied between $0,5,\infty$.
Although all the panels shown in this figure are for pullers
($\beta>0$), very similar results (not shown) are found for pushers
($\beta<0$). Provided the data is represented in this way, we conclude
that the stresslet parameter $\beta$ is relatively unimportant in
determining the clustering and phase behaviour of active particles
with hydrodynamics, and that disklike ``pushers'', ``pullers'' and
``shakers'' should all show similar behaviour. The small value of
$\alpha$ needed to ensure this -- for other values the region of
strong fluctuations shifts in vertical position with changing $\beta$
-- suggests that $B_2$ has a relatively large effect compared with
$B_1$ in determining the overall magnitude of effective activity.
This is consistent with the notion that $B_1$ acts even in the limit
of zero volume fraction whereas $B_2$ affects the activity levels
through interactions only.  Accordingly the effect of $B_2$ should
scale as the inverse particle separation, which is large for the
relatively crowded, clustered regimes of interest here.

Finally, we turn from the properties of the statistically steady state
that forms a long time after the system was randomly initialised at
time $t=0$, to comment briefly on the dynamics of the aggregated
domains that slowly coarsen at low $\va$.  See Fig.~\ref{npctsquirm},
which shows that the number of particles per domain $\bar{n}^*(t)$
appears to coarsen as a power law $t^{2/3}$, corresponding to a
characteristic linear domain radius $t^{1/3}$, which we note is the
same power law as for the coarsening dynamics of an undriven system in
a diffusion dominated regime~\cite{chaikin95}.  Given that the
squirmers have hydrodynamic interactions and ballistic rather than
diffusive intercollisional dynamics, this correspondence is a
nontrivial and perhaps even surprising result.

\section{Conclusions}
\label{sec:conclusions}

Motivated by recent experiments reporting clustering in active
suspensions~\cite{palacci13,PhysRevLett.108.268303,buttinoni13,PhysRevX.5.011004},
we have studied with full hydrodynamics the phase behaviour of active
particles with an interaction potential that has an attractive
Lennard-Jones (LJ) component, together with a steep repulsive
contribution that diverges at particle contact $r\to 2R^+$, where $R$
is the particle radius.

As a preliminary step we mapped out the equilibrium phase diagram of
passive particles subject to the same interaction potential, finding
phase behaviour broadly as expected for a LJ system, with regions of
G-S, G-L and L-S phase coexistence at low temperatures.  However
compared to the LJ potential used more conventionally in the
literature, the repulsive component of which diverges only as $r\to
0$, we found crystalline order to set in a slightly lower packing
fractions, consistent with the greater excluded volume effects from
the additional steeply repulsive contribution to our modified
potential as $r\to 2R$.  This preliminary study also established that
the main features of phase behaviour can indeed be mapped out with
only a small number of particles, $N=242$, which is the maximum
feasible system size in the much more costly hydrodynamic simulations.

As a second preliminary step, we then studied active Brownian
particles (ABPs) subject to the same potential, but without
hydrodynamics. As in previous studies of ABPs with the conventional
Lennard-Jones potential~\cite{redner13b}, we observed two distinct
regimes of phase separation. For low activities a regime of
attraction-induced phase separation is seen, strikingly ressembling
that at low temperatures in the equilibrium phase diagram, with a
return to more homegeneous states for higher activities.  In this
sense, activity acts in a manner somewhat analogous to the temperature
of a passive system.  For ABPs with a small rotational diffusion
coefficient, however, we found at high activities a second regime of
phase separation that has no counterpart in the passive system and is
instead a purely non-equilibrium phenomenon known as motility induced
phase separation (MIPS)~\cite{tailleur2008,cates2010,cates15}.

Motivated by the important effects of hydrodynamics in many active
systems, we proceeded finally to simulate a suspension of hydrodynamic
squirmers subject to the same (modified) LJ potential.  The central
results of this hydrodynamic study are fourfold.

First, we showed that activity again acts in a manner somewhat
analogous to an effective temperature, in the sense that a regime of
equilibrium-like attraction-induced phase separation is seen at low
activities (with the activity suitably adimensionalised by the
strength of the attractive potential), with a return to much more
homogenised states at high activity.

Second, at high values of the (scaled) activity we find only weak
stringlike clustering, with no evidence for motility induced phase
separation (MIPS). By careful comparison with the ABPs, we interpreted
this weak clustering as the signature of a nearby MIPS that has been
largely suppressed by hydrodynamic interactions according to the
arguments in Ref.~\cite{fielding14}.

In view of this suppression, we tentatively suggest that clustering
phenomena seen experimentally in active colloids in which
hydrodynamics is important are more likely to be in the regime of
equilibrium-like attraction-dominated phase separation than the regime
of purely non-equilibrium MIPS. Indeed, as noted above, the presence
of a small phoretic attraction can seldom be ruled out in synthetic
colloids. It is also interesting to recall that clustering is seldom
seen in bacterial
suspensions~\cite{sokolov07,Shawnd13,PhysRevLett.110.228102}, at
least in the absence of an obvious depletion
interaction~\cite{Schwarz-Linek13032012,C0SM00214C,Dorken07122012}.

Third, we showed that the regime of phase separation seen at low
activity, despite being broadly analogous to its equilibrium
counterpart in the passive systems, also shows important differences.
In particular we find a lower than expected degree of hexatic
ordering, compared to the degree of clustering, consistent with the
more dynamic nature of so these ``living
crystals''~\cite{buttinoni13,palacci13,PhysRevX.5.011004}. Activity is
therefore seen to disrupt the level of translational ordering compared
to that in the corresponding equilibrium phase diagram.

Fourth, we showed that an extended definition of the rescaled activity
parameter allows us to recover strikingly similar phase diagrams
across the full range of the stresslet parameter $\beta$, suggesting
that ``pushers'', ``pullers'' and ``shakers'' all behave in a similar
manner with regards their aggregation and phase behaviour.

In future work it would be interesting to consider more fully the
consequences of dimensionality on these effects. Recall that the system
simulated in this work concerns a 2D layer of active particles with 2D
hydrodynamic interactions between them. While we are confident, on the
basis of checks reported earlier in Ref.~\cite{fielding14},
that the same effects would also obtain for a 2D layer of particles
with 3D hydrodynamic interactions, it would clearly be interesting to
consider the case of 3D packings with 3D hydrodynamics.  Indeed, loose
aggregates moving coherently for long times were observed previously
for a 3D suspension of purely repulsive squirmers with 3D
hydrodynamics in Ref.~\cite{Alarcon201356}.  In
Ref.~\cite{PhysRevLett.112.118101} a strongly confined 2D monolayer of
purely repulsive squirmers was observed to show strong phase
separation. However it should be noted that an important effect of
strong confinement will be to screen hydrodynamic interactions. In
this sense, therefore, the observation of strong phase separation in
Ref.~\cite{PhysRevLett.112.118101} is perhaps not surprising.
Far more puzzlingly, phase separation was apparently observed in an
unconfined 2D monolayer of purely repulsive squirmers with fully 3D
hydrodynamics in Ref.~\cite{ishi08b}, though at an area fraction
$\phi=0.1$ that seems far too low for MIPS to be implicated.

Asymmetric particles could be considered in order to mimic chemotactic
swimmers, with their rich phase diagrams
\cite{goles07,saha14,soto14b,yang14}. A closer connection with
experiments on phoretic active colloids might be established by
solving the dynamics of a chemical concentration field in the
fluid between the particles, with a suitable boundary condition at the
particle surfaces, to account for diffusiophoresis.

Finally, in view of the important role played by the ratio $\zeta$ of
the angular reorientation time to the interparticle collision time, it
would be interesting to simulate elongated particles with full
hydrodynamics.  One might anticipate that particle elongation would
significantly affect the angular reorientation time, perhaps leading
to an effective emergent value of $\zeta$, even with hydrodynamics,
that would predispose a return to MIPS at high values of activity.  It
remains an open challenge to determine the link, if any, between that
possible mechanism for MIPS in elongated particles and the one
originally put forward in
Refs.~\cite{toner95,toner98,Toner2005170,ramastoner03,PhysRevE.77.011920,PhysRevLett.101.268101}
in the context of active liquids crystals.

\section{Acknowledgements} The research leading to these results has
received funding from the European Research Council under the European
Union's Seventh Framework Programme (FP7/2007-2013) / ERC grant
agreement number 279365. The authors thank Mike Cates, Mark Miller, Ignacio Pagonabarraga and Francisco Alarc\'on for discussions.

\end{document}